\documentclass[12pt,aps,prd,nofootinbib,superscriptaddress, onecolumn,preprintnumbers,balancelastpage]{revtex4}
\pdfoutput=1
\usepackage{amssymb,amsmath,amsfonts,amstext,graphics, multirow}
\usepackage{graphicx}
\usepackage{epstopdf}
\usepackage{url}
\usepackage{appendix}
\usepackage{color,slashed}
 \definecolor{orange}{cmyk}{0,0.25,1,0}

\newdimen\tableauside\tableauside=1.0ex
\newdimen\tableaurule\tableaurule=0.4pt
\newdimen\tableaustep
\def\phantomhrule#1{\hbox{\vbox to0pt{\hrule height\tableaurule width#1\vss}}}
\def\phantomvrule#1{\vbox{\hbox to0pt{\vrule width\tableaurule height#1\hss}}}
\def\sqr{\vbox{%
  \phantomhrule\tableaustep
  \hbox{\phantomvrule\tableaustep\kern\tableaustep\phantomvrule\tableaustep}%
  \hbox{\vbox{\phantomhrule\tableauside}\kern-\tableaurule}}}
\def\squares#1{\hbox{\count0=#1\noindent\loop\sqr
  \advance\count0 by-1 \ifnum\count0>0\repeat}}
\def\tableau#1{\vcenter{\offinterlineskip
  \tableaustep=\tableauside\advance\tableaustep by-\tableaurule
  \kern\normallineskip\hbox
    {\kern\normallineskip\vbox
      {\gettableau#1 0 }%
     \kern\normallineskip\kern\tableaurule}%
  \kern\normallineskip\kern\tableaurule}}
\def\gettableau#1 {\ifnum#1=0\let\next=\null\else
  \squares{#1}\let\next=\gettableau\fi\next}

\newcommand{\gsim}{\lower.7ex\hbox{$\;\stackrel{\textstyle>}{\sim}\;$}}
\newcommand{\lsim}{\lower.7ex\hbox{$\;\stackrel{\textstyle<}{\sim}\;$}}
\def\OO{{\cal O}}
\def\BB{{\cal B}}
\def\LL{{\cal L}}

\def\EE{{\cal E}}

\def\GG{{\cal G}}
\def\MM{{\cal M}}

\newcommand{\TeV}{\,\mathrm{TeV}}
\newcommand{\GeV}{\,\mathrm{GeV}}
\newcommand{\MeV}{\,\mathrm{MeV}}

\newcommand{\fb}{\,\mathrm{fb}}
\newcommand{\ifb}{\,\mathrm{fb}^{-1}}
\newcommand{\pb}{\,\mathrm{pb}}
\newcommand{\ipb}{\,\mathrm{pb}^{-1}}

\newcommand{\half}{{\frac{1}{2}  }}

\newcommand{\hc}{\text{ h.c. }}

\newcommand{\MET}{ \slashed{E}{}_T}

\newcommand{\be}{\begin{eqnarray}}
\newcommand{\ee}{\end{eqnarray}}
\newcommand{\bea}{\begin{eqnarray}}
\newcommand{\eea}{\end{eqnarray}}

\newcommand{\bef}{\begin{figure}[htbp]\begin{center}}
\newcommand{\eef}{\end{center}\end{figure}}

\newcommand{\go}{\tilde{g}}
\newcommand{\LSP}{{\chi^0}}

\begin{document}
\def\thesection       {\arabic{section}}
\def\thetable       {\arabic{table}}
\def\thefigure       {\arabic{figure}}
\def\thesubsection       {\thesection.\arabic{subsection}}

%%%%%%%%%%%%%%%%%%%%%%%%%%%%%%%%%%%%%%%%%%%%%%%%
\title{Where the Sidewalk Ends:\\Jets and Missing Energy Search Strategies for the 7 TeV LHC}

\author{Daniele S. M. Alves}
\affiliation{Theory Group, SLAC National Accelerator Laboratory, Menlo Park, CA 94025}
\affiliation{Stanford Institute for Theoretical Physics, Stanford University, Stanford, CA 94306}
\author{Eder Izaguirre}
\affiliation{Theory Group, SLAC National Accelerator Laboratory, Menlo Park, CA 94025}
\affiliation{Stanford Institute for Theoretical Physics, Stanford University, Stanford, CA 94306}

%$^{a,b}$ and 
\author{Jay G. Wacker}
\affiliation{Theory Group, SLAC National Accelerator Laboratory, Menlo Park, CA 94025}
\affiliation{Stanford Institute for Theoretical Physics, Stanford University, Stanford, CA 94306}

%$^{a}$}
%%\\
%$^a$ 
%Theory Group,\\ 
%SLAC National Accelerator Laboratory,\\
%Menlo Park, CA 94025\\\\
%$^b$Stanford Institute for Theoretical Physics,\\ 
%Stanford University,\\ 
%Stanford, CA 94306
%}
\begin{abstract}
%\abstract{
 This work explores the potential reach of the 7 TeV LHC to new colored states in the context of simplified models and addresses the issue of which search regions are necessary to cover an extensive set of event topologies and kinematic regimes. This article demonstrates that if searches are designed  to focus on specific regions of phase space, then new physics may be missed if it lies in unexpected corners.
Simple multiregion search strategies can be designed to cover all of kinematic possibilities.  A set of benchmark models are created that cover the qualitatively different signatures and a benchmark multiregion search strategy is presented that covers these models.
%}
\end{abstract}
\maketitle
\newpage
\preprint{}
\tableofcontents

\vspace{0.5in}
\noindent
{\em
There is a place where the sidewalk ends\\
And before the street begins,\\
And there the grass grows soft and white,\\
And there the sun burns crimson bright,\\
And there the moon-bird rests from his flight\\
To cool in the peppermint wind.\\
} -- ``Where The Sidewalk Ends'' Shel Silverstein, 1974

\section{Introduction}
By the end of 2011, the LHC is expected to have accumulated $\OO(1\fb^{-1})$ of integrated luminosity at $\sqrt{s}=7\TeV$. At this early stage, models of new physics with large production cross sections should be prioritized in experimental analyses. Among those are spectra with colored particles that decay into jets and a stable invisible particle, which are a common prediction of extensions of the Standard Model motivated by the hierarchy problem and the dark matter puzzle.

The Tevatron carried out a variety of searches in jets and missing energy and extended bounds on colored objects in specific contexts \cite{Portell:2006qb,Abulencia:2006kk,:2007ww, Wang:2009zzf}. Most searches, however, were optimized for mSUGRA-type benchmarked scenarios, that are affected by strong assumptions on the spectrum, mass splittings and branching ratios, and therefore underrepresent the kinematic possibilities and decay topologies. 
At the LHC, both on the theory and experiment sides, similar model-specific studies for new physics prospects have been carried out  \cite{ Baer:1995nq,  Cheng:2002ab, Macesanu:2002db, Macesanu:2002ew, Falkowski:2005ck, Baer:2006id, Kawagoe:2006sm,Yetkin:2007zz, Baer:2007eh, Cho:2007fg, DeSanctis:2008zz, Ozturk:2009fj, Lungu:2009nh,  Feldman:2009zc, Baer:2009dn,  Kim:2009nq, recent}. Moreover, many analyses are often obscured by the presentation of the results in terms of high energy mSUGRA parameters such as $m_0$ and $m_{1/2}$, making it non-trivial to translate the bounds for alternative theories. Previous work recast the Tevatron bounds for more general scenarios and showed that its reach could have been significantly extended, had a reanalysis been performed with a less benchmark driven and more comprehensive search strategy \cite{Alwall:2008ve, Alwall:2008va}.  Some model-independent searches have been carried at the Tevatron \cite{:2007kp, Aaltonen:2007dg, Henderson:2008ps}, but the inferred limits from these searches have never been performed and are difficult to do {\em a posteriori}.

An alternative paradigm for creating searches and exploring common features of new physics was recently put forward and dubbed ``simplified models.''   Simplified models parameterize the new physics  by a simple particle spectrum, its production mode and decay topologies with the masses, cross sections and branching ratios taken as free parameters \cite{Alwall:2008ve, Alwall:2008va, ArkaniHamed:2007fw,Alwall:2008ag, Izaguirre:2010nj, CERNWorkshop, Alves:2010za}. These simplified models capture generic kinematic properties of models that are relevant for early searches.  Particles that are not involved in a specific signature are decoupled from the simplified model. In some sense simplified models define a perturbation series in model space that approximates specific theories and facilitates  their mapping into experimental observables. 
While still model-dependent, simplified models help reduce model dependence that often plague top-down parameterizations of new physics.
Simplified models have been used in the context of solving the ``LHC Inverse Problem'' \cite{ArkaniHamed:2005px} and also in characterizing experimental anomalies \cite{Alwall:2008ag}.   Other approaches in reducing model dependence of searches have been to more fully explore the 19 parameter ``phenomenological MSSM'' \cite{Berger:2008cq, Conley:2010du}.  

In this article, the early LHC discovery potential for jets and missing energy signatures is studied in the framework of simplified models. The simplified spectrum considered here consists of a gluino-like object that is pair-produced and decays to jets and a stable neutral particle that escapes detection. The decay can proceed directly to the neutral particle or through a cascade, in which case the spectrum is augmented by intermediate particles. The decay modes were chosen so as to cover a diverse kinematic range and correspond to two-body direct decays,  three-body direct decays, three types of one-step cascades and a two-step cascade. The scope of these simplified models is broader than the realm of supersymmetric theories \cite{Plehn:2008ae, Baer:2010uy, Appelquist:2000nn, Cheng:2002iz, Cheng:2003ju,Gregoire:2008mr} as long as spin correlations are irrelevant in the discovery process \cite{Nojiri:2011qn}.

The first result of this article is a systematic  quantification of the reach for these colored objects with an extensive set of combined cuts on $\MET$ (missing energy) and $H_T$ (visible energy).  These results are used to subsequently extract the optimal sensitivity of early LHC searches to the gluino branching ratios times production cross section. 
For a QCD reference cross section and 100\% gluino branching ratio to jets plus $\MET$, the LHC may find a $2\sigma$ evidence for gluinos as heavy as 800 GeV in models with large mass splittings, with $1 \ifb$ of integrated luminosity. For compressed spectra, the estimated current limit of 150~GeV on gluino masses can be extended up to  500 GeV.

The optimal reach for the set of simplified models considered is not obtained by a single set of cuts on any set of variables, but instead requires searching in several distinct regions.  The second result of this article presents a minimal set of search regions  designed to obtain a near-optimal reach  for each simplified model.
Six search regions are necessary to cover the entire mass parameter space of the simplified models, guaranteeing that signatures of new physics in all hadronic jets + MET channels will not be missed. These six searches are:
\begin{itemize}
\item {\bf dijet high MET} for  compressed spectra
\item {\bf trijet high MET} for heavy gluinos decaying via 2-body direct decay to the LSP
\item {\bf multijet low MET} for light gluinos decaying via 3-body direct decay or cascade decay to the LSP
\item {\bf multijet  moderate MET} for intermediate mass  gluinos decaying via 3-body direct decay or cascade decay to the LSP
\item {\bf multijet  high MET} for heavy gluinos decaying via 3-body direct decay or cascade decay to the LSP
\item {\bf multijet high $\mathbf{H_T}$} for heavy gluinos decaying to light LSP's.
\end{itemize}

The third result of this article is  a short list of benchmark simplified models that can be used to ensure that all regions of parameter space are being covered with equal diligence.  These benchmark models are useful because they provide examples of spectra  that are not represented in mSUGRA or other existing parameterizations. 

The outline of this article is as follows.  Sec.~\ref{Sec: Models} provides a precise definition of the simplified models used in this study. Sec.~\ref{MC} describes the Monte Carlo (MC) implementation of signal and background. These MC calculations are used on Sec.~\ref{cross-sections} for the estimation of the cross section sensitivities, and on Sec.~\ref{cuts} for investigating and designing a comprehensive search strategy on $\MET$ and $H_T$.
Sec.~\ref{Sec: MultipleDecayModes} explores how to use the results of single topology simplified models to multiple topology simplified models.
The conclusions are summarized on Sec.~\ref{discussion}.
Finally, App.~\ref{App: PlotHell} shows the estimated reach for the simplified models with $45\ipb$ and $1\ifb$ and App.~\ref{App: Benchmarks} gives a set of fully specified benchmark simplified models.

\section{Simplified Models for Colored Octets}
\label{Sec: Models}

The simplified models described in this section are effective field theories after electroweak symmetry breaking.  Therefore the states of the theory will have well-defined quantum numbers under $SU(3)_c\times U(1)_\text{EM}$. 
The two principle states in the set of simplified models are a color octet Majorana fermion, $\go$, and a stable neutral Majorana fermion, $\LSP$.  The topologies considered in this article will all commence with $\go$ being pair produced through its QCD interactions.     
The fully differential production cross section of $\go$ is assumed to be proportional to the fully differential tree-level QCD production cross section.  This is a good approximation whenever non-Standard Model states do not play a dominant role in the production of $\go$.
Important cases such as resonant production are not directly proportional to the differential QCD production cross section and deserve a separate treatment.
The results derived in Sec.~\ref{cross-sections} and Sec.~\ref{cuts} are parametrized in terms of $m_{\go}$ and $m_{\LSP}$ and the product of the  inclusive production cross section of $\go$ and any relevant branching ratios.  
By allowing the total $\go$ production to be rescaled, many different theories are effectively described by a single simplified model.
Effects such as different $\go$ multiplicities or spin degrees-of-freedom and even $t$-channel squarks in supersymmetric theories are well-captured by this parameterization of simplified models.
For convenience, the inclusive production rates will be referenced  against  the QCD next-to-leading order cross section.   
 
The decay of each $\go$ into $\LSP$ and additional jets proceeds through an effective operator. 
The most straight-forward augmentation of this simplified model is to add intermediate particles in 
 the decay chain of $\go$ which can significantly alter the decay kinematics. Two benchmark particle types are chosen in this study: a charged Dirac fermion $\chi^\pm$ with the quantum numbers of a chargino and a neutral Majorana fermion ${\chi'}^0$ with the quantum numbers of a neutralino.  In addition to the direct decay of
\begin{eqnarray*}
\go ~\rightarrow~ \LSP+X
\end{eqnarray*} 
 where $X= g \text{ or } q\bar{q}$,
 two other decay routes are studied: a one-step cascade decay,
 \begin{eqnarray*}
 \go~\rightarrow~ \chi^\pm+X~\rightarrow~ \LSP+W^\pm+X, 
 \end{eqnarray*}
 and a two step cascade decay, 
 \begin{eqnarray*}
\go~\rightarrow~ \chi^\pm + X~\rightarrow~ \chi'{}^0+W^\pm+X~ \rightarrow ~\LSP +Z^0+W^\pm+X.
\end{eqnarray*} 

In the cascade decays of the $\go$ down to $\LSP$, leptons may appear as by-products of the $W^\pm$ and $Z^0$ decays. 
Although leptons may be present in the final state, this study will be concerned exclusively with all hadronic final states plus missing energy. 
Leptonic channels can in principle increase the sensitivity to simplified models with electroweak gauge bosons in the final state; however, these decay topologies are ``lepton poor'' due to the relatively small branching fraction of electroweak vector bosons into leptons. In this study, the sensitivity to $\go$ one-step decays through a chargino in the one-lepton channel was found to be worse than that in the all hadronic mode. 

The four $\go$ decay modes considered are detailed below and illustrated in
Fig.~\ref{Fig: Spectra}.

\subsection{Two-body direct decay}

\begin{center} {\large $\go\rightarrow g \LSP$} \end{center}

In this simplified model, the  dominant decay mode is $\go\rightarrow \LSP + g$.  This decay proceeds through an effective operator of the form
\begin{eqnarray}
\LL_{\text{int}} = \frac{ 1}{M} \go \sigma^{\mu\nu} \LSP G_{\mu\nu} +\hc
\end{eqnarray}
This decay mode can occur in gauge mediated supersymmetry breaking models when the gluino is the next-to-LSP and $\LSP$ is the gravitino.   
Two body $\go$ decays to $\LSP$ and a gluon also appear at loop level in standard supersymmetric theories and can become large when there are sizeable splittings between the left and right handed squarks and can even become the dominant decay mode in split supersymmetric models \cite{Gluino2Body, RabyOlogy}.  

In Universal Extra-Dimensions (UED), the KK-excitation of the gluon may dominantly decay to a gluon and a KK-graviton \cite{Appelquist:2000nn}. Note that in this case the pair-produced octet is a massive spin-one particle and the invisible final state is a massive spin-2 particle. Nevertheless, such UED scenarios can still be parametrized by the simplified model considered here given that the inclusive production cross section is taken to be a free parameter and most of the signature's features are determined by kinematics alone.  

In addition to theories where the colored particle is a color octet, this simplified model also well-approximates theories where the new colored particle lies in a different representation but have the same decay topology,
such as squark NLSPs.   The approximation breaks down when initial or final state radiation becomes important.

\begin{figure}[t]
\begin{center}
\includegraphics[width=6in]{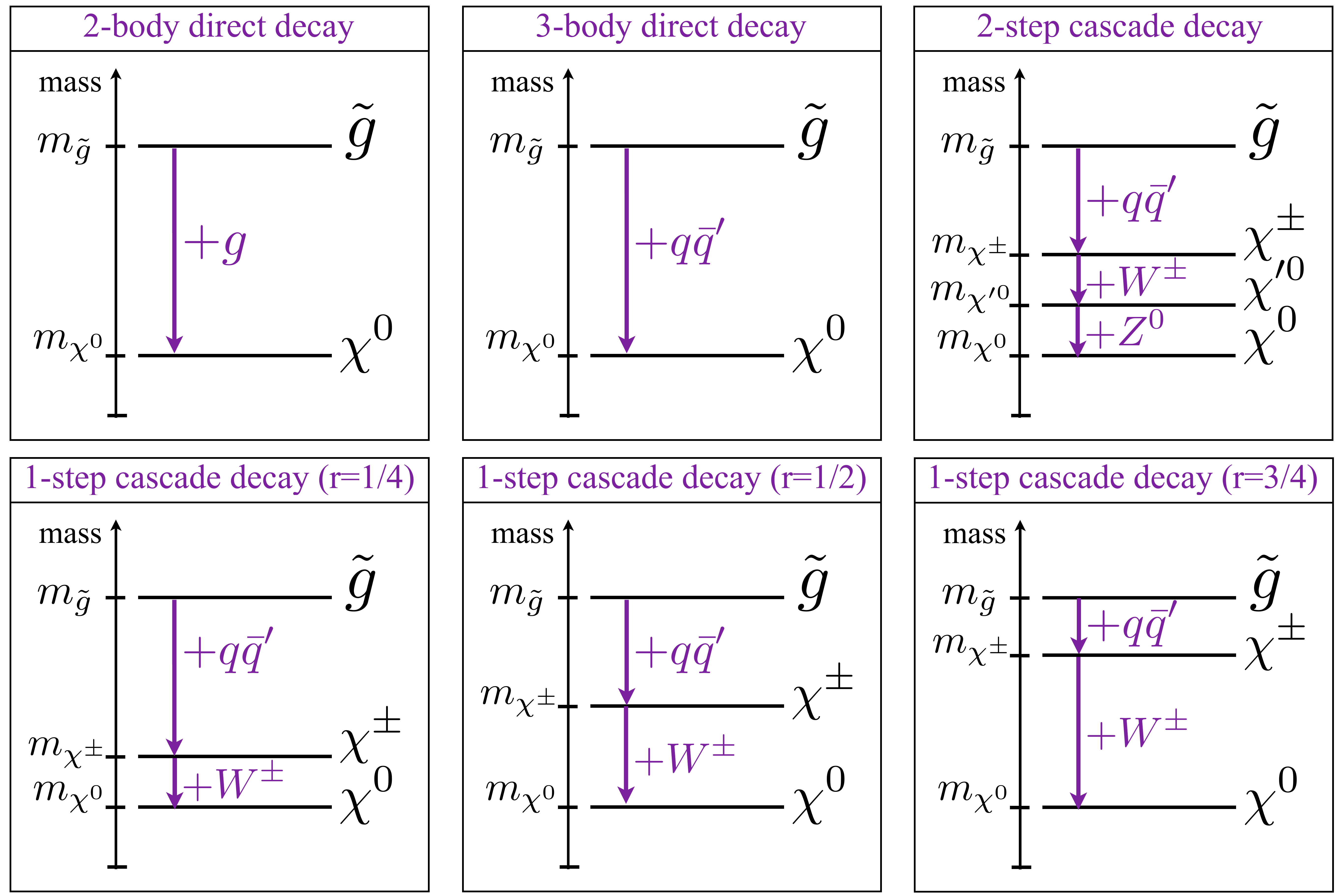}
 \end{center}
 \caption{
  \label{Fig: Spectra}
 Simplified spectra and decay modes considered in this study. The masses of the pair produced colored state, $\go$, and of the the invisible final state, $\LSP$, are varied over the range from 100 GeV to 1 TeV. The masses of the intermediate states in the decay chains are held at a fixed distance from $m_{\go}$ and $m_{\LSP}$, corresponding to slices through the mass parameter space that are representative of all the kinematical configurations.}
 \end{figure}

\subsection{Three-body direct decay}
\begin{center}{\large $\go\rightarrow q \bar{q}^\prime \LSP$}
\end{center}

In contrast to the previous simplified model, the gluino decay may still be direct, but proceed via 3-body. In supersymmetric models with decoupled squarks, {\it e.g.} in split-supersymmetry \cite{Kilian:2004uj}, the gluino can decay to an electroweak gaugino and two light flavored quarks. 
This decay proceeds through a dimension-6 operator
\begin{eqnarray}
\LL_{\text{int}} =   \frac{\tilde{g}_{i \chi}^2}{M_{i}^2} \go q_i \bar{q}_i\LSP +\hc .  
\end{eqnarray}
where $i$ runs over the different quark flavors and $\tilde{g}_{i \chi}$ is the Yukawa coupling between the quark-squark and and $\LSP$.  This article is assuming that the decays proceed into light flavored quarks.    The final state flavor structure is determined by the mass spectrum of the corresponding squarks with the decay through lighter mass squarks occurring more rapidly.   
Assuming all electroweakinos are kinematically accessible  in gluino decays, direct three body decays dominate in the following cases
\begin{itemize}
\item $\LSP = \tilde{B}$  and the right handed squarks are lightest,
\item $\LSP = \tilde{W}$ and the left handed squarks are lightest,
\item $\LSP = \tilde{H}$ and the heavy flavor quarks are kinematically accessible in gluino decays .
\end{itemize}
In AMSB scenarios, the LSP is usually wino-like and there no strong hierarchy in squark masses; however, due to the fact that the wino gauge-Yukawa coupling is larger, the direct decay of the gluino has a large branching ratio.
In mSUGRA and GMSB-like models, the LSP is usually bino-like and there is no strong splitting between the left and right-handed squarks; therefore, the direct decays usually do not dominate.  
The study of heavy flavor decays are beyond the scope of this article, see  \cite{HeavyFlavor}.

This simplified model is relevant not only for 3-body direct decays, but it also effectively describes cascade decays in which there is a large mass splitting between the gluino and the first intermediate state in the cascade and the rest of intermediate states are compressed near the LSP.
This occurs frequently when the LSP is an $SU(2)_L$ multiplet such as the wino or Higgsino.   In these scenarios,  the mass splittings between the LSP and NLSP are typically less than $50\GeV$ and can be as small as $\OO(100 \MeV)$.

\subsection{One-step cascade decay}

\begin{center}{\large $\go\rightarrow q \bar{q}^\prime {\chi}^\pm \rightarrow q \bar{q}^\prime (W^\pm \LSP)$}
\end{center}

In this simplified model, the gluino goes through a 3-body direct decay to a chargino that subsequently decays to a gauge boson and the LSP. This simplified model is commonly realized in mSUGRA, and more generally the chain gluino $\rightarrow$ heavy electroweakino $\rightarrow$ lightest electroweakino is preferred in many supersymmetric scenarios \cite{Barnett:1987kn}. A similar chain KK-gluon $\rightarrow$ KK-gauge boson $\rightarrow$ KK-graviton is also present in Extra Dimensions, because spin correlation are  a mild effect on early discovery.

One of the challenges in non-minimal simplified models is the proliferation of parameters.   There are two additional parameters to consider  in one-step cascade decays: the mass of the intermediate particle $m_{\chi^\pm}$ and the branching ratio of $\go$ decaying into $\chi^\pm$.
Each simplified model will initially be considered with branching ratios set to 100\%.
Multiple decay modes can be studied by taking linear combinations of single decay modes.  This is discussed in Sec.~\ref{Sec: MultipleDecayModes}.

By setting branching ratios to 100\%, the number of parameters in one step cascades are reduced to  four, namely $m_{\go}$, $m_{\LSP}$, $m_{\chi^\pm}$, and $\sigma(pp\rightarrow \go\go+X)$.  The choice of $m_{\chi^\pm}$ alters the kinematics of the theory and therefore all four parameters are important.  This article explores this four dimensional parameter space by choosing several distinct relationships between the three masses of the theory (or ``mass slices''). 
The mass slices  were chosen to maximize the distinctive features of one-step cascades and capture all the relevant corners of phase space. Those are specified by the following choice of  intermediate chargino masses:
\begin{equation}
\label{Eq: 1stepSlicings}
m_{{\chi}^\pm}=m_{\LSP}+r(m_{\go}-m_{\LSP}),
\end{equation}
where three values of the parameter $r$ chosen are: $r=$1/4, 1/2 and 3/4.
The case of $r=0$ is identical to the direct 3-body decay.  The case of $r=1$ closely resembles the direct two-body decay when the boost of the $W^\pm$ becomes large and its decay products merge together.  For the mass ranges of interest for the LHC, the $W^\pm$ is a light particle and therefore most theories are in this boosted $W^\pm$ regime.
   
An alternative mass slice that could be studied is
\begin{eqnarray}
\label{Eq: CharginoThreshold}
m_{{\chi}^\pm}\simeq m_{\LSP}+ m_{W^\pm}
\end{eqnarray}
in order to explore the effects of on-shell decays near threshold.    This mass slicing is present in a subspace of the three parameterizations adopted in Eq.~\ref{Eq: 1stepSlicings} when 
\begin{eqnarray}
m_{\LSP} = m_{\go} -m_{W^\pm}/r .
\end{eqnarray}
Threshold effects are fairly modest because the mass scales accessible at the LHC are sufficiently above $m_{W^\pm}$. 
The results presented in Sec.~\ref{Sec: Results} and App.~\ref{App: PlotHell} do not show anomalous behavior near this line.   
Threshold effects are important for lighter $\go$ masses.  In \cite{Alves:2010za}, this can be seen as a sharp drop in the cross section sensitivity along the line in Eq.~\ref{Eq: CharginoThreshold}.

Given that the intermediate particle in this simplified model is a chargino, all events have  two $W^\pm$ bosons in the final state. 
 Clearly, alternative simplified models resembling this one exist, in which the intermediate state is  neutral and decays to a $Z^0$ boson  instead of a $W^\pm$. 
One of the open questions in the study of simplified models is how adjacent theories  in ``model space'' approximate each other.   This example of exchanging a neutralino for a chargino, or equivalently a final state $W^\pm$ for a $Z^0$, is a prime case study for gaining intuition for the scope of simplified models.  

When exchanging a $W^\pm$ for a $Z^0$, the mass difference is a small effect at the LHC.  The primary difference is the presence of leptonic events which get vetoed in the searches considered in this study.  
Ultimately, there are two questions that need to be addressed. 
The first is whether the optimization of search regions is affected by the choice of final state vector bosons.  This question is the most critical because it can lead to a delay in the discovery of new physics.
Fortunately, the search design is insensitive to the small difference in the number of leptonic events between the two theories.

The second question is how to translate the limits from one simplified model to the other.   Answering this question requires understanding the differences in the acceptances/efficiencies for events with $Z^0$-final states versus $W^\pm$-final states.   
%The cut efficiencies do not differ significantly for most mass spectra, 
The efficiencies are similar for most mass spectra, with larger discrepancies arising  in regions of heavy $\go$ and light $\LSP$, differing by typically 20\% due to the $20 \GeV$ lepton veto.
This results in a slight gain in sensitivity for simplified models with $Z^0$ bosons in the final-state.  
To understand the magnitude of this difference, consider the following back-of-the-envelope calculation.   
The lepton veto is not perfectly efficient in identifying leptons.
The efficiency for finding isolated individual leptons in the jet-rich environment of $\go$ decays is estimated by 
{\tt PGS4} to be
\begin{eqnarray}
\epsilon_{\ell} \simeq 74\% 
\end{eqnarray}
with approximately equal efficiencies for both electrons and muons.
This causes a leakage of leptonic events into the signal region. 
The events in this sample have two $W^\pm$ bosons with branching ratios 
\begin{eqnarray*}
\begin{array}{rclcccr}
B_{2W_{0\ell}} &\equiv&\text{Br }( 2 W\rightarrow \text{hadrons}) &\simeq&  B_{W_h}^2& \simeq &62\% \\
B_{2W_{1\ell}} &\equiv&\text{Br }( 2 W\rightarrow \ell+ \text{hadrons})&\simeq& 2 B_{W_h}(1- B_{W_h})&\simeq & 33.5\% \\
B_{2W_{2\ell}} &\equiv&\text{Br }( 2 W \rightarrow 2 \ell)& \simeq& (1- B_{W_h})^2 &\simeq &4.5\%.
\end{array}
\end{eqnarray*}
where
\begin{eqnarray*}
B_{W_h}=  \text{Br }(W\rightarrow q\bar{q}') + \text{Br }(W\rightarrow \tau \nu)\simeq 82.4\% .
\end{eqnarray*}
This results in 
a  leakage fraction of leptonic events into the all-hadronic channel of
\begin{eqnarray}
L_{2W} \equiv (1-\epsilon_\ell) B_{2W_{1\ell}} + (1-\epsilon_\ell)^2B_{2W_{2\ell}}\simeq  9\%.
\end{eqnarray}
Thus the resulting fraction of $2 W^\pm$ events passing the lepton veto is
\begin{eqnarray}
\epsilon_{2W}^{\text{ $\ell$-veto}} \simeq B_{2W_{0\ell}} + L_{2W} \simeq 71\%.
\end{eqnarray}
The analogous reasoning for events with $2Z^0$ in the final state leads to  
\begin{eqnarray}
\epsilon_{2Z}^{\text{ $\ell$-veto}}\simeq 88\%  \qquad \Rightarrow\qquad \epsilon_{2Z}^{\text{ $\ell$-veto}}/\epsilon_{2W}^{\text{ $\ell$-veto}} \simeq 1.23.
\end{eqnarray}
%of the events have both Z's going hadronically, invisibly or to $\tau^+\tau^-$, 12.5\% of the events contain two final-state leptons and the remaining 0.5\% contain 4 final-state leptons. Hence, a fraction of $0.26^2\times0.125+0.26^4\times0.005$ of the events leak into the hadronic sample, resulting in a 88\% lepton veto efficiency for events with a pair of $Z$'s.
 This back-of-the-envelope calculation is surprisingly accurate for the large mass splitting regions and accounts for the $20\%$ difference  in sensitivity mentioned above for $W^\pm$ versus $Z^0$ simplified models. For more compressed spectra where the $W^\pm$/$Z^0$'s do not have phase space to go on-shell, the final-state leptons are so soft that the leptonic ``leakage" rates into the hadronic region can easily go up to 100\%. In such cases there is virtually no distinction between the two types of simplified models, implying nearly identical search sensitivities.

\subsection{ Two-step cascade decay}

\begin{center} {\large $\go\rightarrow q \bar{q}^\prime {\chi}^\pm \rightarrow q \bar{q}^\prime (W^\pm {\chi}'{}^0) \rightarrow q \bar{q}^\prime (W^\pm (Z^0 \LSP))$}
\end{center}

The last simplified model in this study 
consists of $\go$ decaying through two intermediate states to $\LSP$.
These are encountered in the MSSM if the ordering of the electroweakinos is $\tilde{W}-\tilde{H}- \tilde{B}$ or $\tilde{B}-\tilde{H}- \tilde{W}$ because the decay widths of $\tilde{W}\leftrightarrow \tilde{B}$ are suppressed relative to the $\tilde{B}\leftrightarrow \tilde{H} \leftrightarrow \tilde{W}$ transitions.
In extensions to the MSSM with singlets such as the NMSSM,
the decay chain $\go \rightarrow \tilde{W} \rightarrow \tilde{B} \rightarrow \tilde{S}$ is a common cascade because the singlino, $\tilde{S}$, can have a small coupling to the MSSM. 

Higher-step cascades are possible, but they typically have small branching ratios due to phase space suppression and the sheer number of combinatoric possibilities.
Furthermore, the kinematics frequently resemble the simplified models already considered.

The number of parameters in this two-step cascade grows considerably over the three in the direct decay simplified models and five in one step cascade decays.  In two-step cascade decays there are  four masses, three branching ratios and the production cross section.  As with the one-step cascade decay, the branching ratio for the two-step cascade decay is set to 100\%.  Sec.~\ref{Sec: MultipleDecayModes} considers going away from this limit.

In order to simplify the four dimensional mass parameter space,
the following mass slicing is chosen for  the intermediate states
\begin{eqnarray}
\nonumber
m_{\chi^\pm}&=&m_{\LSP}+r(m_{\go}-m_{\LSP}) \\
m_{\chi'{}^0}&=&m_{\LSP}+r'(m_{{\chi}^\pm}-m_{\LSP}),
\end{eqnarray}
with $r=r'=1/2$.  More choices could be studied, but this specific parameterization maximizes the difference in the kinematics between the two-step cascade decay and the one-step and direct decay simplified models previously considered.

The effects of the $W^\pm$ and $Z^0$ going on-shell in the cascade decay occur for the following values
\begin{eqnarray}
\nonumber
m_{\LSP} = m_{\go} - m_{Z^0}/r r' &\qquad& \text{ $Z^0$ on-shell threshold}\\
m_{\LSP} = m_{\go} - m_{W^\pm}/r(1- r') &\qquad& \text{ $W^\pm$ on-shell threshold} .
\end{eqnarray}
Only one of these can be satisfied at a time unless  $r' =1/(1+ m_{W^\pm}/m_{Z^0})\simeq 0.53$ which is close to the value chosen.    The threshold effect can be seen in the results in Sec.~\ref{Sec: Results} and in App.~\ref{App: PlotHell} as a loss of sensitivity near the line $m_{\go} \simeq m_{\LSP} + 350 \GeV$.

This simplified model has  $\go$ decaying into a charged ${\chi}^\pm$.  $\chi^\pm$ subsequently decays to a $W^\pm$ boson and a neutral ${\chi}'{}^0$.  Finally, ${\chi}'{}^0$  decays to a $Z^0$ boson and $\LSP$. The final state  therefore contains four jets,  two $W$'s,  two $Z^0$'s, and two $\LSP$.  
Alternate charge choices for the intermediate states can be made and result in different combinations of vector bosons in the final state such as those containing $4 W^\pm$ or $4Z^0$.
As in the one-step cascade decay, the lepton veto is responsible for the dominant differences between the sensitivity to these final states.
A similar estimation can be made for  the differences between $2W^\pm 2Z^0$ final states versus $4W^\pm$ or $4Z^0$ final states. 
The resulting search sensitivity to $2W^\pm2Z^0$ simplified models is typically 16\% worse than for 4$Z^0$ simplified models and 16\% better than for 4$W^\pm$ simplified models.

\section{Backgrounds and Signal Simulation}
\label{MC}

The dominant Standard Model backgrounds to jets and $\MET$ signatures are $t\bar{t}+ \text{ jets}$, $W^{\pm}+ \text{ jets}$, $Z^0+ \text{ jets}$, and QCD. In this study, subdominant processes such as single top and diboson production in association with jets are not included.  The matrix elements for parton level events are computed in \texttt{MadGraph 4.4.32} \cite{Alwall:2007st} with  CTEQ6L1  parton distribution functions used throughout \cite{Pumplin:2002vw}.   Variable renormalization and factorization scales are set to the transverse energy of the event \cite{Bauer:2009km}. The SM parton level processes generated are 
\begin{eqnarray}
\label{Eq: Samples}
pp  \rightarrow W^{\pm} + nj  &\quad& 1\le n\le 3\\
\nonumber
pp  \rightarrow Z^{0} + nj &\quad & 1\le n \le 3 \\
\nonumber
pp  \rightarrow t\bar{t} + nj &\quad & 0\le n \le 2\\
\nonumber
pp \rightarrow W^{\pm}+b\bar{b}+nj &\quad & 0\le n \le 2\\
\nonumber
pp \rightarrow Z^0+b\bar{b}+nj &\quad & 0\le n \le 2
\end{eqnarray}
where $j$ stands for gluons and light flavored quarks.
The $Z^0$ and $W^{\pm}$ plus jets samples are forced to decay to final states involving neutrinos, but no restrictions on $t\bar{t}$ events are placed. 

The Standard Model contribution to missing energy distributions peaks at low energies, whereas many signatures of new physics yield events with extremely high missing energy.   Therefore it is important to have sufficient Monte Carlo statistics on the tail of the $\MET$ distribution.  
To achieve sufficient statistics, different samples are generated for each SM process, where each sample has the $p_T$ of the massive particle lying in a given interval.
For instance, $Z^{0}+\text{jets}$ parton level events are broken into three samples with
%\footnote{For a description on the generation of the rest of the SM backgrounds, see {\tt http://www.lhcbackgrounds.com}}:
%%
\begin{eqnarray*}
0~\GeV\le &p_{T,Z^0}&\le200\GeV\\
200~\GeV < &p_{T,Z^0}&\le300\GeV\\
300~\GeV< &p_{T,Z^0}&.
\end{eqnarray*}
 In the case of $t\bar{t}$, where there are two heavy particles, the samples are divided by the larger $p_{T}$ of either top quark in the event. 

Contributions from QCD to jets and $\MET$  can come from either detector effects and jet energy mismeasurement, or neutrinos appearing in the decay of heavy flavor hadrons. To estimate the QCD contribution to jets and $\MET$ signatures the following subprocesses are generated
\begin{eqnarray}
\nonumber
pp\rightarrow jjjj& &\\
pp\rightarrow b\bar{b}+nj &\quad & 0\le n \le 2
\end{eqnarray}
The QCD $4j$ background is calculated by subdividing the samples by the $p_T$ of the two leading jets in the event at parton level. Eight separate samples are generated, specified in Fig.~\ref{Fig: Jet Generation}.
\begin{figure}
\begin{center}
\includegraphics[width=3in]{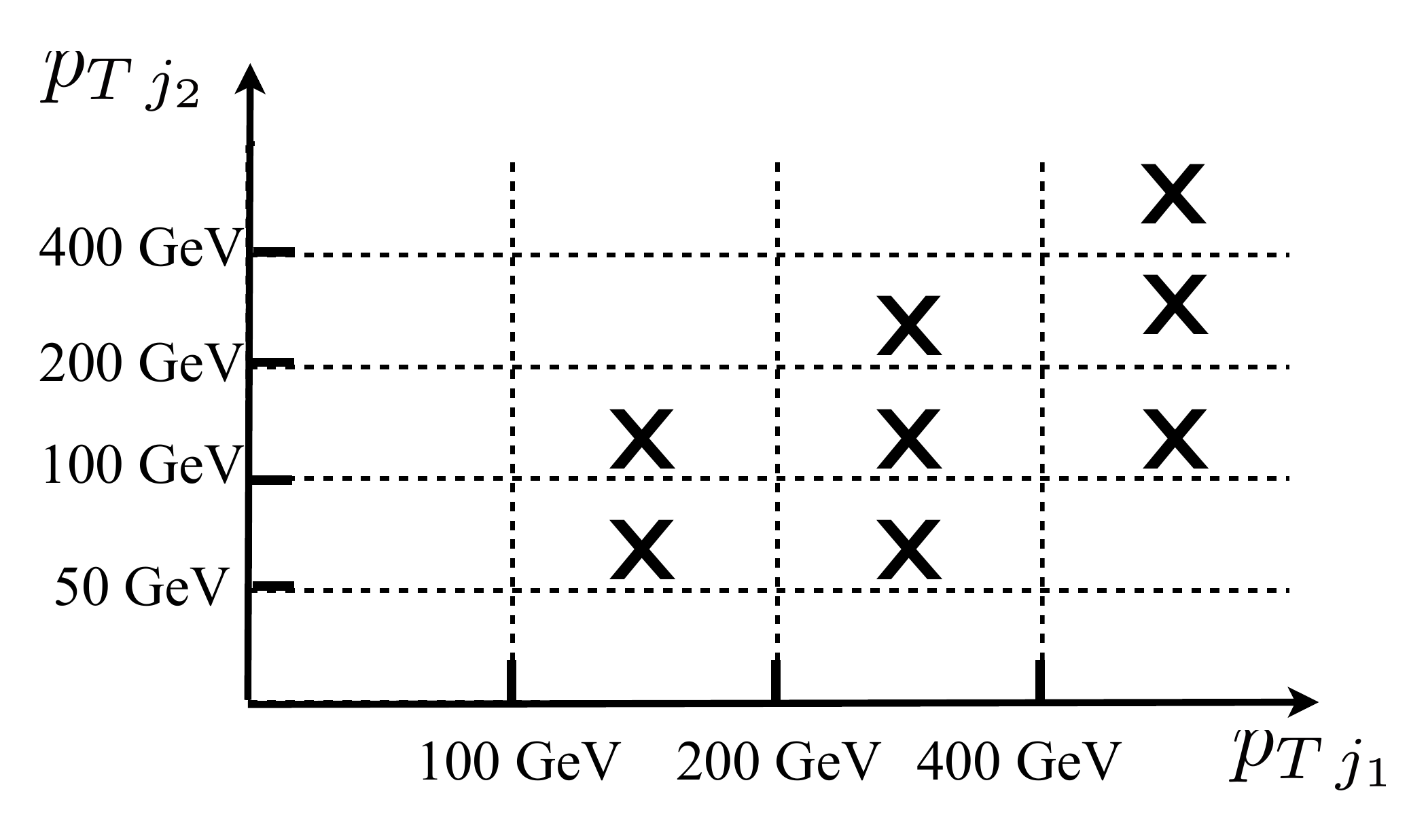}
\caption{\label{Fig: Jet Generation} Schematic representation of the different samples generated for QCD 4j processes. The events in a given sample have the $p_T$ of the two hardest jets within a chosen range.  The lower right box can not be populated the four jet events because $p_{T\; j_1} > p_{T\; j_2}+p_{T\; j_3}+p_{T\; j_4}$. }
\end{center}
\end{figure}

The signal, $\go$ pair-production, was generated in association with up to two jets at parton level,
\begin{eqnarray}
pp\rightarrow \go\go + nj \quad 0\le n \le 2.
\end{eqnarray}
The effects of including additional radiation in signal processes have been documented in several studies \cite{Alwall:2008ve,Alwall:2008va,BSMMatching}.

For both signal and backgrounds, the showering, hadronization, particle decays, and matching of parton showers to matrix element partons were done in \texttt{PYTHIA 6.4} \cite{Sjostrand:2006za}.  
%In this study, a $k_{\perp}$ jet-clustering algorithm is used for identifying the jets within {\tt PYTHIA}.  
Matching of parton showers to matrix elements is an important step to correctly utilize the multiple final state samples generated above.  The matrix elements better describe hard radiation, while the parton shower generates softer radiation that fills out jets \cite{PSME}.  The MLM
 parton shower/matrix element matching scheme is used with a shower-$k_{\perp}$ scheme introduced in \cite{SUSYPSME}. The matching scales used in this article are listed in the table below.
 \begin{eqnarray}
\begin{array}{|c|c|}
\hline
\text{Sample}& Q_{\text{Match}}\\
\hline\hline
t\bar{t}+\text{jets}& 100 \GeV\\
V+ \text{jets}& 40\GeV\\
V+ b\bar{b}+\text{jets}& 40\GeV\\
b\bar{b}+\text{jets}& 50\GeV\\
\go\go +\text{jets}& 100\GeV\\
\hline
\end{array}
 \end{eqnarray}
 %%
 
  %Given the universality of initial and final state radiation accompanying colored particle production,  multijet events with missing energy can become important in setting the strongest limits. 
  %The signal and tops plus jets backgrounds were matched up to two jets, whereas electroweak backgrounds were matched up to three jets. 
 Hard jets  beyond the multiplicities listed in \eqref{Eq: Samples} must be generated by the parton shower.    In particular, for $W^\pm +\text{ jets}$ and $Z^0+\text{ jets}$, the fourth jet and beyond are generated through the parton shower.
 This approximation has been validated by several studies.
 For instance, in $W^{+}+\text{jets}$, the discrepancy in the inclusive rate for four jets and $\MET$ from matching up to three jets versus four jets is $\OO(15\%)$ \cite{PSME}.

Next-to-leading-order (NLO) corrections alter the predictions of both signal and background.  With parton shower/matrix element matching, the shapes of differential distributions are accurately described by tree level predictions.  The largest corrections are to the inclusive production cross section and can be absorbed in $K$-factors. The leading order cross sections of the signal are normalized to the NLO cross sections calculated in \texttt{Prospino 2.0} \cite{Beenakker:1996ch}. The $t\bar{t}+X$ leading order production cross section is scaled to the NLO one with the same K factor used in \cite{Izaguirre:2010nj} obtained from \cite{Kidonakis:2008mu}. As a cross-check, the $t\bar{t}+$jets $\MET$ distribution is checked to agree with that from \cite{AtlasCuts}. Similarly, the $W/Z+$ jets leading order cross sections are scaled with the same normalization from \cite{Berger:2009ep, Maitre:2009xp, Campbell:2003hd}, and checked to agree with \cite{AtlasCuts}.

 \texttt{PGS 4} is used as a ``transfer function'' that takes hadron level events  to  reconstructed, detector-level objects that represent the objects that experiments at the LHC report results on  \cite{PGS}.  This study uses the {\tt PGS 4}  ATLAS card and this has been shown to reproduce results to $\OO(20\%)$ accuracy.   
 
 One of the drawbacks of {\tt PGS 4} is that it uses  a cone jet algorithm with $\Delta R =0.7$.  This is an infrared unsafe jet algorithm, but better represents the anti-$k_T$ algorithms used by the experiments than the $k_T$ algorithm.   
 The Standard Model backgrounds change by at most $\OO(10\%)$
 when varying the cone size to $\Delta R=0.4$ .

 The signal offers a more varied testing ground for the effects of changing the jet algorithm.   Two competing effects are found. The first is that there is more out-of-cone energy  for smaller cones, resulting in less energetic jets.  The second effect is that smaller cone jet algorithms find more jets.   These two effects are more pronounced when contrasting  a jet-poor 2-body decaying $\go$ with a jet-rich 2-step cascade decaying $\go$. 
 The  dependence of the kinematic cut efficiencies on $\Delta R$ varies with mass splitting between the $\go$  and $\LSP$.  For compressed spectra, when the $p_T$ of the jets is reduced,  the efficiencies for the smaller cone size decrease because jets fall below the minimum jet $p_T$ requirement. 
 For widely spaced spectra, where jets are energetic, more jets are found with a smaller cone size and the efficiency to have multiple jets passing the minimum jet requirement increases.  Altogether, the efficiencies differ by at most $\OO(20\%)$ and is consistent with other studies \cite{Krohn:2009zg} .
 This effect is not included in the remainder of this article, where the analysis is performed with a fixed $\Delta R=0.7$.

\section{Optimal Sensitivities}
\label{cross-sections}
\label{Sec: Results}

One of the goals of this article is to explore the necessary kinematic cuts that will optimally distinguish signal from background.
%The simulated background and signal events, scanned over a large region of the $m_{\go} - m_{\LSP}$ parameter space with varied decay modes, were used for an extensive exploration of kinematic cuts. 
The simplified models described in Sec.~\ref{Sec: Models} provide a varied signal space when the full range of $m_{\go}$ and $m_{\LSP}$ is considered.  
The first step is setting which kinematic variables will be investigated.  Several options were explored:
\begin{itemize}
\item  missing transverse energy, $\MET$,
\item  visible transverse energy, $H_T$,
\item  transverse momentum of the leading jets, $p_{Ti}$ with $i=1,2,3,4$,
\item   effective mass, $M_\text{eff}$,\footnote{$M_\text{eff}$ is defined as $M_\text{eff} \equiv \mbox{$E_T\hspace{-0.22in}\not\hspace{0.13in}$} ~+\sum\limits_{i=1}^{n}p_{Ti}$, where $n$ is the channel's jet multiplicity.}
\item fractional missing energy, $\MET/M_\text{eff}$,
\item fractional jet momenta, {\it e.g.} $p_{T1}/\MET$ .
\end{itemize}
Combinations of these kinematic variables were considered in defining search regions in conjunction with varying multiplicities of jets.   Since the primary concern of this article is the discovery of new states, as opposed to measurement of their properties, the most effective strategy is to keep the largest fraction of the signal possible.  Placing cuts on several variables, as in a multivariate analysis, has the drawback that it increases the complexity of the search.  
In this article, search regions are defined by at most two kinematic cuts applied after predefined selection criteria.
These selection criteria, motivated by typical LHC searches for jets and missing energy in all-hadronic channels, are:
\begin{itemize}
\item  a jet $p_T$ selection of $p_{T1}\geq100\GeV$ and $p_{Ti}\geq50\GeV$, for $2\leq i \leq\text{channel multiplicity}$;
\item a  $p_{T\ell} >20\GeV$ veto on all isolated leptons; 
\item  a $|\Delta\phi (j_i, \MET)| >0.2$ cut between that transverse momentum vector of the $i^{\text{th}}$ jet and the missing transverse momentum vector, where $i=1, 2, 3$ \footnote{For the dijet channel, this cut is only applied to the leading two jets.};
\item a $|\eta_{j_i} |<2.5$ requirement on the momentum vector of the $i^{\text{th}}$ jet, where $i=1,2,3$\footnote{For the dijet and trijet channels, this cut is only applied to the leading two or three jets, respectively.};
\item a $\MET/M_\text{eff}>0.30, 0.25, 0.20$ cut for dijet, trijet and multijet channels, respectively,  for controlling QCD backgrounds.
\end{itemize}

The expected $2\sigma$-sensitivity quantifies how well a search region performs for a simplified model at a given point  in the $m_{\go}$, $m_{\LSP}$ and $\sigma_\text{prod}$ parameter space.
A given model is declared to be within the 2$\sigma$-reach of a search region if the number of signal events passing the search region cuts is equal or greater than twice the background uncertainties, both statistical and systematic:
\begin{equation}
\label{Eq: MinEvents}
N_{\text{signal}}\geq\Delta B^{2\sigma} = 2\times\sqrt{(\Delta B_\text{stat})^2+(\Delta B_{\text{syst}})^2},
\end{equation}
where a 30\% systematic background uncertainty, $\Delta B_{\text{syst}} = 0.30 N_{\text{bkg}}$, is assumed throughout this article. The statistical error on the number of background events, $\Delta B_\text{stat}$, is the Poisson fluctuation of the expected number of background events and is equal to $\sqrt{N_\text{bkg}}$ in the Gaussian limit \cite{PDG}.
For a given search region, the minimum value of $\sigma(pp\rightarrow\go\go)\times\BB(\go\rightarrow nj+\MET)$ that satisfies relation \eqref{Eq: MinEvents} can  be computed
\begin{eqnarray}
\label{Eq: SigmaLimit}
\sigma_{\text{prod}}^{2\sigma}(pp\rightarrow\go\go X)\times\BB(\go\rightarrow nj+\MET) = \frac{\Delta B^{2\sigma}}{A\times\epsilon\times\LL},
\end{eqnarray}
where $A$ is the acceptance and $\epsilon$ is the efficiency of the search region's cuts for the signal.

The reach of the LHC depends on the choice of search regions. In order to extract the optimal reach for each point in parameter space, a large ensemble of search regions in the kinematic variables discussed above was explored. In the final outcome, it turned out that combined cuts in $\{\MET, H_T\}$, optimized individually for each simplified model, yielded the best sensitivity to the production cross section times branching ratio of the parameter space explored. In practice, that was obtained by simulating simplified models in a grid of $\{m_{\go},m_{\LSP}\}$ for all six decay modes considered, in 50~GeV steps for $m_{\go}$ and 25~GeV steps for $m_{\LSP}$. Each simulated simplified model had its reach evaluated in a grid of combined cuts in $\{\MET, H_T\}$, with $100\GeV\leq\MET\leq600\GeV$ in 50~GeV steps and $\MET\leq H_T\leq1200\GeV$ in 50~GeV steps.

The optimal sensitivity on the cross section times branching ratio for all mass spectra and decay modes is displayed in Fig.~\ref{CrossSections} and App.~\ref{App: PlotHell}.

% The inclusive multijet ($4^+j$) channel was the most effective on most of the parameter space, with the exception of the degenerate regions, where lower multiplicity channels were found to be more sensitive, given that the visibility of such events rely on the presence of initial and final state radiation \cite{Alwall:2008ve, Alwall:2008va, Izaguirre:2010nj}.

\begin{figure}[ht]
\begin{center}
\includegraphics[width=6.5in]{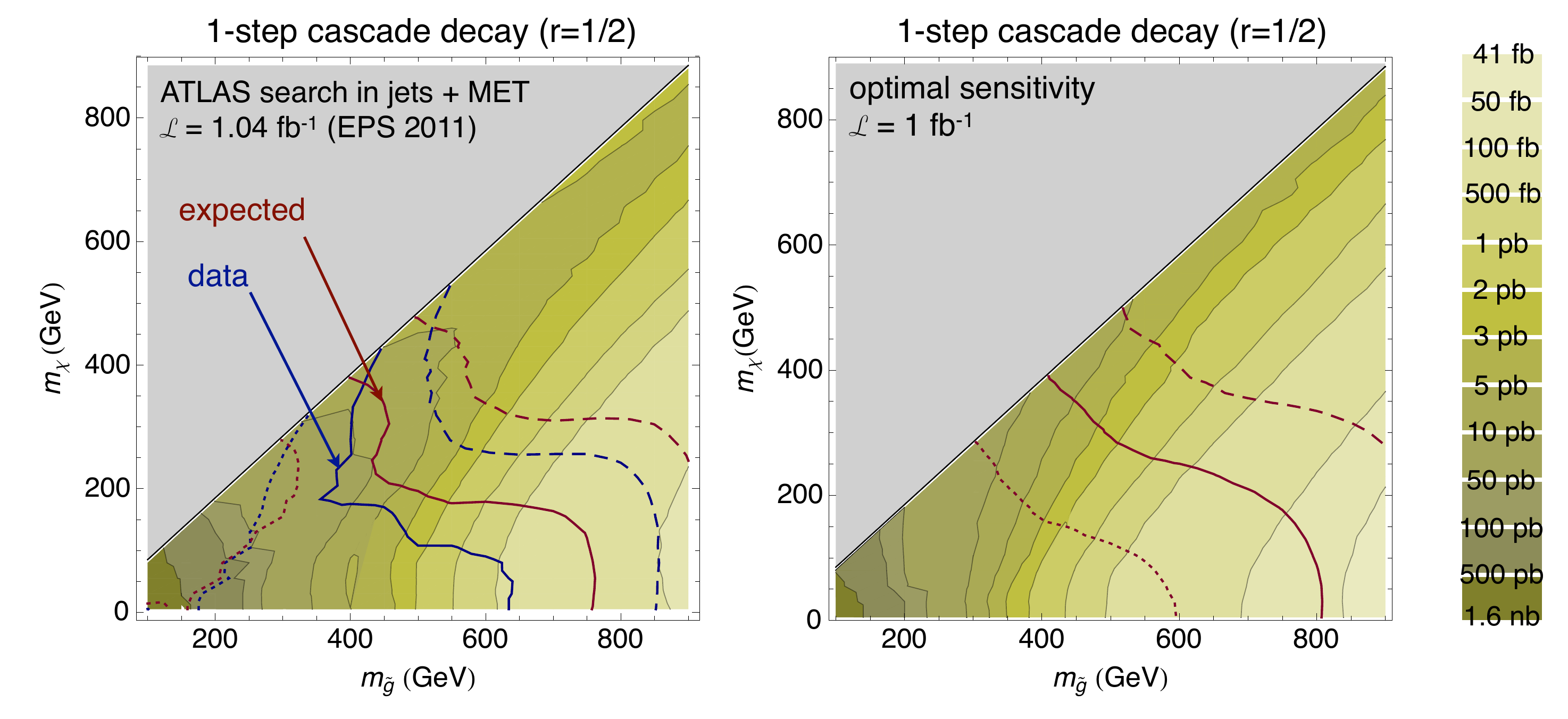}
 \end{center}
 \caption{
  \label{CrossSections}
 Contours of the $2\sigma$-sensitivity for $\sigma(pp\rightarrow\go\go)\times\BB(\go\rightarrow q\bar{q}^\prime\tilde{\chi}^\pm\rightarrow q\bar{q}^\prime (W^\pm\tilde{\chi}))$, with $\LL=1\text{fb}^{-1}$ at 7~TeV.  The expression corresponds to the $\go$ pair-production cross section times the branching ratio for the one step cascade decay mode, where the intermediate particle mass lies at $m_{\chi^\pm}=m_{\LSP}+1/2(m_{\go}-m_{\LSP})$. The optimal reach (right) is contrasted with the reach provided by the $1.04~\text{fb}^{-1}$ ATLAS search for supersymmetry in jets plus missing energy presented at EPS 2011 \cite{ATLAS1.04/fb} (left). The contour values are specified on the color scale on the right. 
 The red lines delimit the expected reach of models for which $\sigma_\text{prod}\times\BB_{\go}$ is a simple parametrization of  the $\go$ NLO-QCD production. They correspond to: (i) $\sigma_\text{prod}\times\BB_{\go}=3\times\sigma_\text{NLO-QCD}$ (dashed line), (ii) $\sigma_\text{prod}\times\BB_{\go}=\sigma_\text{NLO-QCD}$ (solid line), and (iii) $\sigma_\text{prod}\times\BB_{\go}=0.3\times\sigma_\text{NLO-QCD}$ (dotted line). The blue lines in the left plot are analogous to the red lines, but correspond to the estimated $2\sigma$-exclusion regions from the data. 
 }
 \end{figure}

Fig.~\ref{CrossSections} contrasts the optimal $2\sigma$-sensitivities at 7~TeV with the reach of the $1.04~\text{fb}^{-1}$ ATLAS search for supersymmetry in jets plus missing energy presented at EPS 2011 \cite{ATLAS1.04/fb}. The contours specify the minimum accessible values of the $\go$ production cross section times branching ratio, $\sigma_\text{prod}\times\BB_{\go}$, as a function of $m_{\go}$ and $m_{\LSP}$, for the one-step cascade decay with  $r=1/2$. Analogous plots displaying the optimal reach for the other decay topologies considered in this paper are contained in App.~\ref{App: PlotHell} for two luminosity scenarios, $\LL=40\pb^{-1}$ and $1\fb^{-1}$.

The search regions used in the analysis in \cite{ATLAS1.04/fb} are schematically displayed the table below:

\begin{eqnarray*}
\begin{array}{|c|c c c c|}
\hline
\text{Search Region of \cite{ATLAS1.04/fb}}&\geq 2~\text{jets}&\geq 3~\text{jets}&\geq 4~\text{jets}&\text{High Mass}\\
\hline\hline
\MET (\GeV)&>130&>130&>130&>130\\
p_{T1}(\GeV)&>130&>130&>130&>130\\
\hline
p_{T2}(\GeV)&>40&>40&>40&>80\\
p_{T3}(\GeV)& - &>40&>40&>80\\
p_{T4}(\GeV)& - &-&>40&>80\\
\hline
\Delta\phi(p_T,\MET)&>0.4&>0.4&>0.4&>0.4\\
\MET /M_\text{eff}&>0.3&>0.25&>0.25&>0.2\\
M_\text{eff}(\GeV)&>1000&>1000&>500/1000&>1100\\
\hline
\text{Background}&62.3^{\pm4.3}_{\pm9.2}&55^{\pm3.8}_{\pm7.3}&984^{\pm39}_{\pm145}/33.4^{\pm2.9}_{\pm6.3}&13.2^{\pm1.9}_{\pm2.6}\\
\hline
\text{Data}&58&59&1118/40&18\\
\hline
\text{excluded}~\epsilon\times\sigma(\fb)&24&30&477/32&17\\
\hline
\end{array}
\end{eqnarray*}

Fig.~\ref{CrossSections} also shows  the difference in mass reach when optimized cuts are used versus the cuts used in \cite{ATLAS1.04/fb} for a few rescalings  of $\sigma_\text{prod}\times\BB_{\go}$ in terms of the $\go$ NLO-QCD production cross section. Given that for the analysis of \cite{ATLAS1.04/fb} the actual data and background estimates were available, both the expected $2\sigma$-sensitivity and the estimated $2\sigma$-exclusion regions are displayed (red and blue lines, respectively). The difference between expected and estimated exclusion regions is due to a 1-sigma excess in the measured number of events in the multijet signal regions relative to estimated backgrounds. 

Inspection of the sensitivity contours in Fig.~\ref{CrossSections} reveals that, although the ATLAS search regions \cite{ATLAS1.04/fb} were well optimized for heavy spectra and large mass splittings, it compromised its sensitivity for spectra with  $m_{\go}\lsim700\GeV$ and $m_{\LSP}\lsim300\GeV$. One may naively argue that this region has already been excluded and therefore there is no reason for optimizing search sensitivities to $m_{\go}\lsim700\GeV$. That statement is true, however, under the hypothesis that the gluino production cross section is purely determined by its QCD interactions {\it and} that its branching ratio for this 1-step cascade decay mode is equal to $100\%$. As discussed in Sec.~\ref{Sec: Models}, this may not be the case. The production cross section for the gluino may be suppressed relative to the reference QCD cross section through t-channel squark interference, or the gluino may have a significant branching ratio to invisible or soft modes, in which case these lighter spectra remain experimentally viable.

Given that there is no particular reason to expect that CMSSM type-spectra are more likely than lighter or more compressed spectra, it is clear that the LHC discovery potential on jets plus MET searches may be crippled by a limited choice of search regions. On the other hand, measurements of Standard Model backgrounds is a challenging task that may render impractical those analyses demanding several dozens of search regions. The viability of a search strategy that is sensitive to all space of signatures will be addressed in the next section, where a thorough investigation of the $\MET$ and $H_T$ cut-space will be performed,  and guidelines for a comprehensive search strategy will be proposed.

\section{Comprehensive Search Strategies}
\label{cuts}

This section addresses the challenge of creating effective and comprehensive search strategies, which is
how to discover any theory from the entire space of models with the minimum amount of integrated luminosity and a practical number of search regions.
There is no single search region whose coverage is close to optimal in the whole parameter space of theories. On the other hand, there is a practical limitation to the number of search regions that can be studied on a given analysis, due mainly to the challenge posed by measuring the Standard Model background in each search region.  A useful approach to this problem that has been regularly implemented is to design searches that optimize their sensitivity to the regions of parameter space that are theoretically motivated.   Of course, theoretical motivation is subjective and time dependent.   This study adopts a different perspective, namely, to remain agnostic about the likelihood of a given range of model parameters and look for a minimal, comprehensive set of search regions whose combined reach is close to optimal in the full model space. This approach will be referred to as a {\it Multiregion Search Strategy} in jets plus missing energy at the LHC.

\begin{figure}[h]
\begin{center}
\includegraphics[width=6.2in]{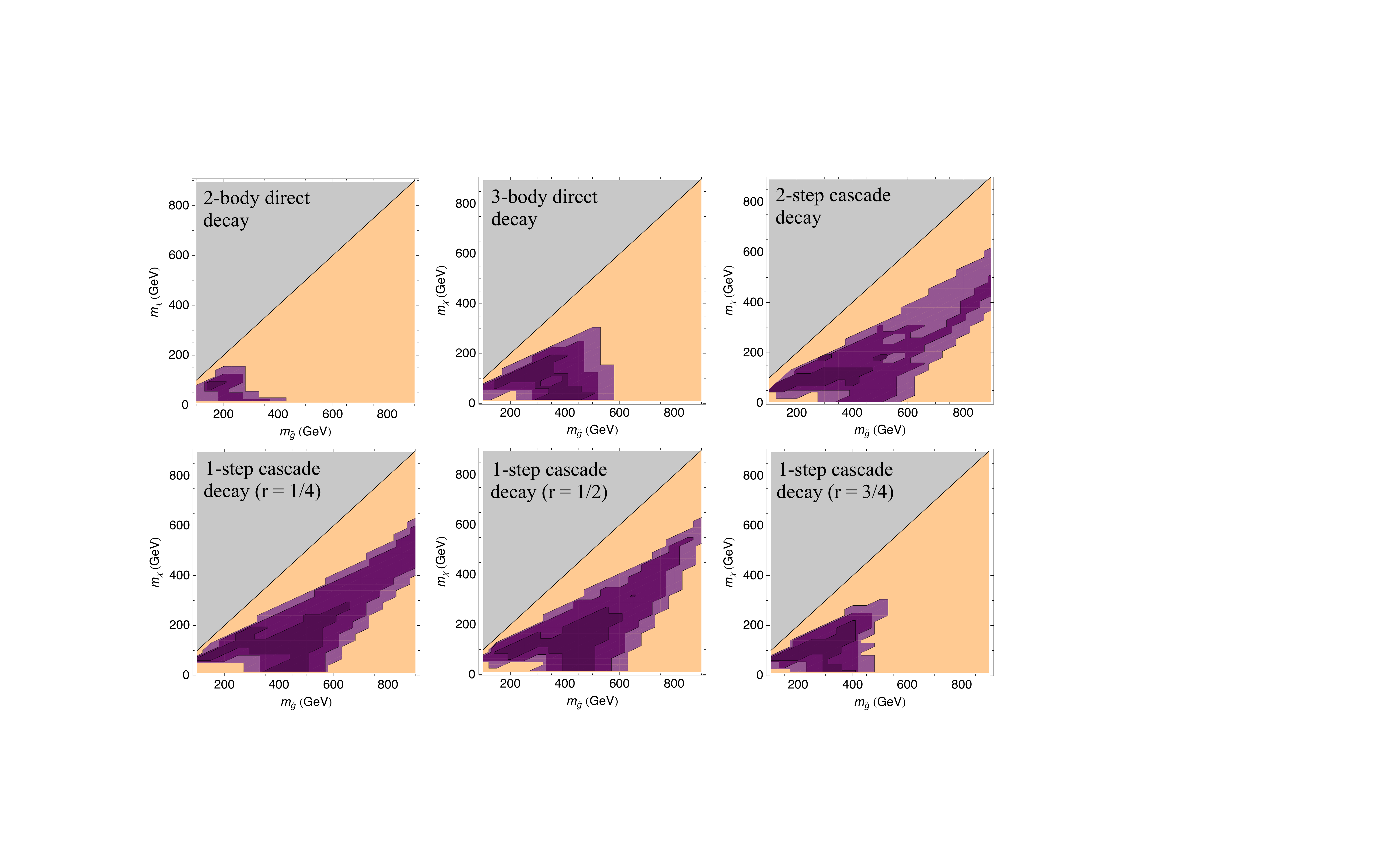}
 \end{center}
 \caption{
  \label{reach}
 Reach of the search region $\mathcal{S}=\{\MET\geq150\GeV$, $H_T\geq750\GeV\}$ in the inclusive multijet channel for the six different decay topologies. The efficacy of this search region is $\EE \le1.10$ in the dark purple region, $\EE \le 1.20$\ in the medium purple region and $\EE \le 1.30$ in the light purple region, assuming an integrated luminosity of $1\text{fb}^{-1}$.}
 \end{figure}

As mentioned in the previous section, an extensive set of search regions were considered in order quantify the optimal sensitivity to our simplified models, characterized by cuts on missing and visible energy within the following range: $100\GeV\leq\MET\leq 600\GeV$ and $\MET\leq H_T\leq1200\GeV$. 
Alternative two-cut search regions, {\it e.g.} $\MET$ and $M_{\text{eff}}$, were investigated as well but were not found to be superior to $\MET$ and $H_T$.
Each search region is efficient in a portion of the model parameter space.  
Conversely, each point in the parameter space may be covered by several search regions.  The goal is to find a minimal set of search regions that covers the entire model parameter space.

{\it Efficacy}, $\EE$, is a variable that quantifies how close to optimal  the reach of a given search region $ \mathcal{S}$ is.  Efficacy is defined as
\begin{equation}
\label{effic}
\mathcal{E}(\mathcal{S}; \mathcal{M})\equiv \frac{\sigma_{\text{prod}} (\mathcal{S}; \mathcal{M})}{
\sigma_{\text{prod}}^{\text{opt}}(\mathcal{M})},
\end{equation}
where $\MM$ denotes a specific simplified model with a given mass spectrum and decay topology, $\mathcal{S}$ is the given search region (specified by a set of cuts), $\sigma_{\text{prod}}(\mathcal{S}; \mathcal{M})$ is the estimated reach in the production cross section for $\MM$ using the search $\mathcal{S}$.  
$\sigma_{\text{prod}}^{\text{opt}}(\mathcal{M})$ is the optimal estimated reach in the production cross section for $\MM$ when considering all search regions,
\begin{eqnarray}
\sigma_{\text{prod}}^{\text{opt}}(\mathcal{M}) \equiv \text{min }\left\{ \sigma_{\text{prod}}(\mathcal{S}; \mathcal{M}) | \mathcal{S}\right\} .
\end{eqnarray}
By definition, $\EE$ is greater than or equal to unity for any search region.   The closer $\EE(\mathcal{S}, \MM)$ is to unity, the more sensitive $\mathcal{S}$ is to $\MM$.  Choosing search regions that have $\EE \simeq 1$ decreases the amount of integrated luminosity necessary to discover new physics.   The goal of this article is to find a set of search regions that covers all possible models with  $\EE < \EE_{\text{crit}}$.
To accomplish this, $\EE(\mathcal{S}; \MM)$ must be computed over the entire space of $\mathcal{S}$ and $\MM$.   As an illustration, Fig.~\ref{reach} shows the reach of the search region $\mathcal{S}=\{\MET\geq150\GeV$, $H_T\geq750\GeV\}$ in the inclusive tetrajet channel for efficacies $\EE_{\text{crit}}=1.1$, $\EE_{\text{crit}}=1.2$ and $\EE_{\text{crit}}=1.3$.
The next step is to find a minimal set of search regions whose combined reach spans all the simplified models and decay topologies under consideration.

\subsection{Multiregion Search Strategy in $\MET$ $\&$ $H_T$}
\label{MSRS}

The search region optimization problem is computationally intensive and has no unique solution. The approach adopted in this study was to perform the optimization with a genetic algorithm. A random ``population" of multiple search regions is initially generated. Mutations and performance selection of the elements of the population are successively implemented, until one or more multiple search regions  with the desired criteria are found.

One of the multisearch regions found by our genetic algorithm with a global efficacy requirement of $\EE_{\text{crit}} = 1.3$ is displayed in Fig.~\ref{MinCuts}. The optimization was performed for three simultaneous values of integrated luminosity: $\LL=10~\text{pb}^{-1},~100~\text{pb}^{-1},~1~\text{fb}^{-1}$. It was found that six search regions were necessary to cover the full model space with $\EE_{\text{crit}} = 1.3$. The search regions for this particular solution are: 
\begin{eqnarray*}
\begin{array}{|c||c||c|c||c|c|c||c|}
\hline
\mathcal{S}&\text{Ch.}& \MET(\GeV)& H_T(\GeV)&W^\pm+nj (\text{fb}) &Z^0+nj (\text{fb}) & t\bar{t}+nj (\text{fb})& \text{Total} (\text{fb})\\
\hline\hline
1&2^+j&>500 &>750&4.8&8.6&0.7&14.1\\
2&3^+j&>450& >500&7.8&12.5&1.8&22.1\\
3&4^+j&>100&>450&182.0&80.9&400.9&667.5\\
4&4^+j&>150&>950&4.3&2.5&6.6&13.4\\
5&4^+j&>250&>300&37.4&29.1&39.1&105.7\\
6&4^+j&>350&>600&6.5&6.1&5.5&18.1\\
\hline
\end{array}
\end{eqnarray*}
where the  background cross section is broken down into its main contributions ( $W^\pm+j$, $Z^0+nj$, $t\bar{t}+j$).   The QCD contribution to $\MET$ drops rapidly. At $\MET \ge 100 \GeV$, the cross section is $\OO(120\fb)$ and by $\MET \ge 150 \GeV$ the cross section has dropped to $\OO(3.5 \fb)$.  Only search region 3 had any appreciable QCD contribution to  the background and was estimated to be $\sim 3.7 \fb$.    Detector effects contributing to $\MET$, modeled by {\tt PGS4}, have large uncertainties, frequently arising from reducible backgrounds.   Therefore the QCD contribution to these search regions is not quoted, but is believed to be subdominant to the listed backgrounds in all search regions.

The minimum number of search regions depends on the global optimality requirement, {\it i.e.}, on $\EE_{\text{crit}}$. For $\EE_{\text{crit}}=1.2$, the number of search regions increases from 6 to 13. If $\EE_{\text{crit}}$ is relaxed to 1.5, a set of 4 search regions  is sufficient. Each search region captures a specific kinematic regime, although the values of the $\MET$ and $H_T$ cuts have some room for variation without affecting the coverage. 
The qualitative kinematic regimes of each search region and the patches of phase space covered by each one of them are described below.

\begin{itemize}
\item {\bf Dijet high MET} 

A high $\MET$ cut on the inclusive dijet channel is required to provide coverage of theories with nearly degenerate spectra \cite{Alwall:2008ve, Alwall:2008va, Giudice:2010wb}.
This qualitative region is expected because compressed spectra are only visible in $\go$ events  that come with initial or final state radiation.   The dijet channel was also important in the heavy $\go$ - light $\LSP$ region for the two-body direct decay mode, where the events typically have two hard jets and very energetic $\LSP$'s.

\item{\bf Trijet high MET} 

The trijet inclusive channel is important for providing sensitivity to heavy $\go$'s directly decaying via 2-body as well.  In contrast to the dijet channel, the trijet is more sensitive to slightly heavier $\LSP$ masses.

\item{\bf Multijet low MET} 

All the other regions were best covered by cuts on the inclusive tetrajet channel. A search region with a low $\MET$ cut is necessary to provide sensitivity to light $\go$'s, specially for the decays that yield high jet multiplicities.

\item{\bf Multijet high $H_T$} 

A high visible energy cut was essential for coverage of models with heavy $\go$'s and large mass splittings in their spectra. For those types of models, the heavy $\go$ decays yield events where a large fraction of the energy going into visible states, allowing the cuts to go far in the distribution tails where backgrounds die off rapidly.

\item{\bf Multijet moderate MET}

A moderate $\MET$ cut is useful for intermediate $m_{\go}$ in regions of moderate mass splittings. It is also important for light $\go$'s decaying via 2-body because this decay mode typically yields fewer but harder jets and more missing energy in comparison to 3-body or cascade decays.

\item{\bf Multijet high MET} 

Finally, a high $\MET$ search region plays an important role for models with heavy $\go$'s decaying to non-relativistic $\LSP$'s. In these types of spectra, $\LSP$ carries off a large fraction of the energy available in the decay, resulting in lower $H_T$ but still a sizable amount of $\MET$.
\end{itemize}

\begin{figure}[t]
\begin{center}
\includegraphics[width=6in]{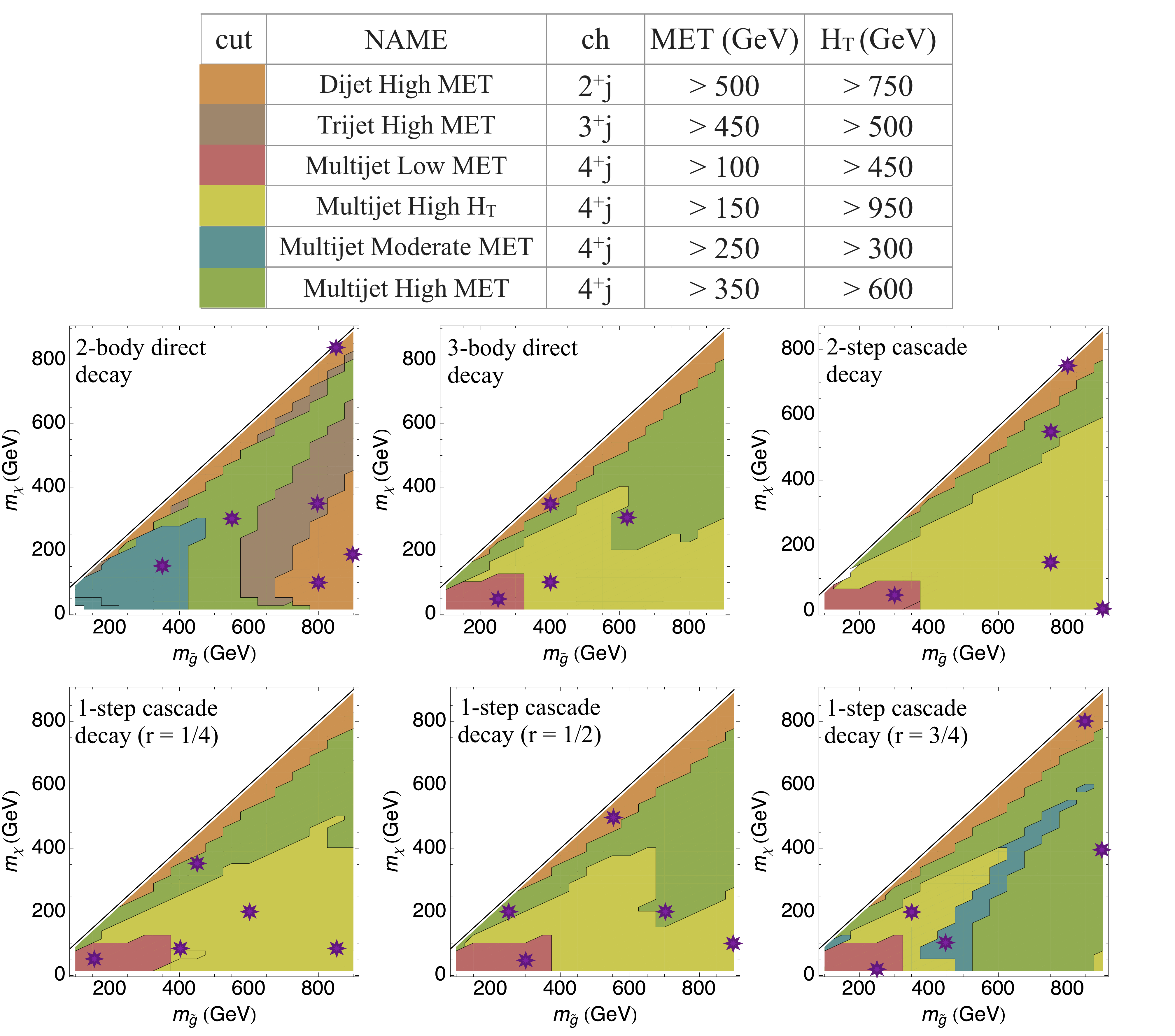}
 \end{center}
 \caption{
  \label{MinCuts}
 Minimal multiple search region on $\MET$ and $H_T$ whose combined reach is within 30\% of optimal for all kinematical regions and decay topologies for the integrated luminosity range $\LL=10\text{pb}^{-1} - 1\text{fb}^{-1}$. The dark dots correspond to benchmark simplified models that are representative of the full phase-space and can be used in optimizing searches, see Tab.~\ref{BM0.1}-\ref{BM2.1}. }
 \end{figure}

Fig.~\ref{HThisto} illustrates the complementarity of the multiple search regions and the limitations of an isolated search region. It displays the $H_T$ distribution in the multijet channel for all the dominant backgrounds and two signal points:

\begin{itemize}
{\item $m_{\go}=200\GeV$, with a 25\% branching ratio into the 1-step cascade decay mode with $r=1/2$ and $m_{\LSP}=55\GeV$ and the remaining decays invisible,}
{\item $m_{\go}=800\GeV$, with a 100\% branching ratio into the 1-step cascade decay mode with $r=1/2$ and $m_{\LSP}=55\GeV$.}
\end{itemize}

The Multijet Low MET search region would cut on $H_T$ around $\sim450\GeV$, making the $m_{\go}=200\GeV$ signal discoverable, but swamping the $m_{\go}=800\GeV$ signal in backgrounds. In order to make the latter signal visible, a hard $H_T$ cut, $\gsim1000\GeV$, is required, as provided by a Multijet High $H_T$ search region, which would however completely kill the $m_{\go}=200\GeV$ signal.

\begin{figure}[t]
\begin{center}
\includegraphics[width=6in]{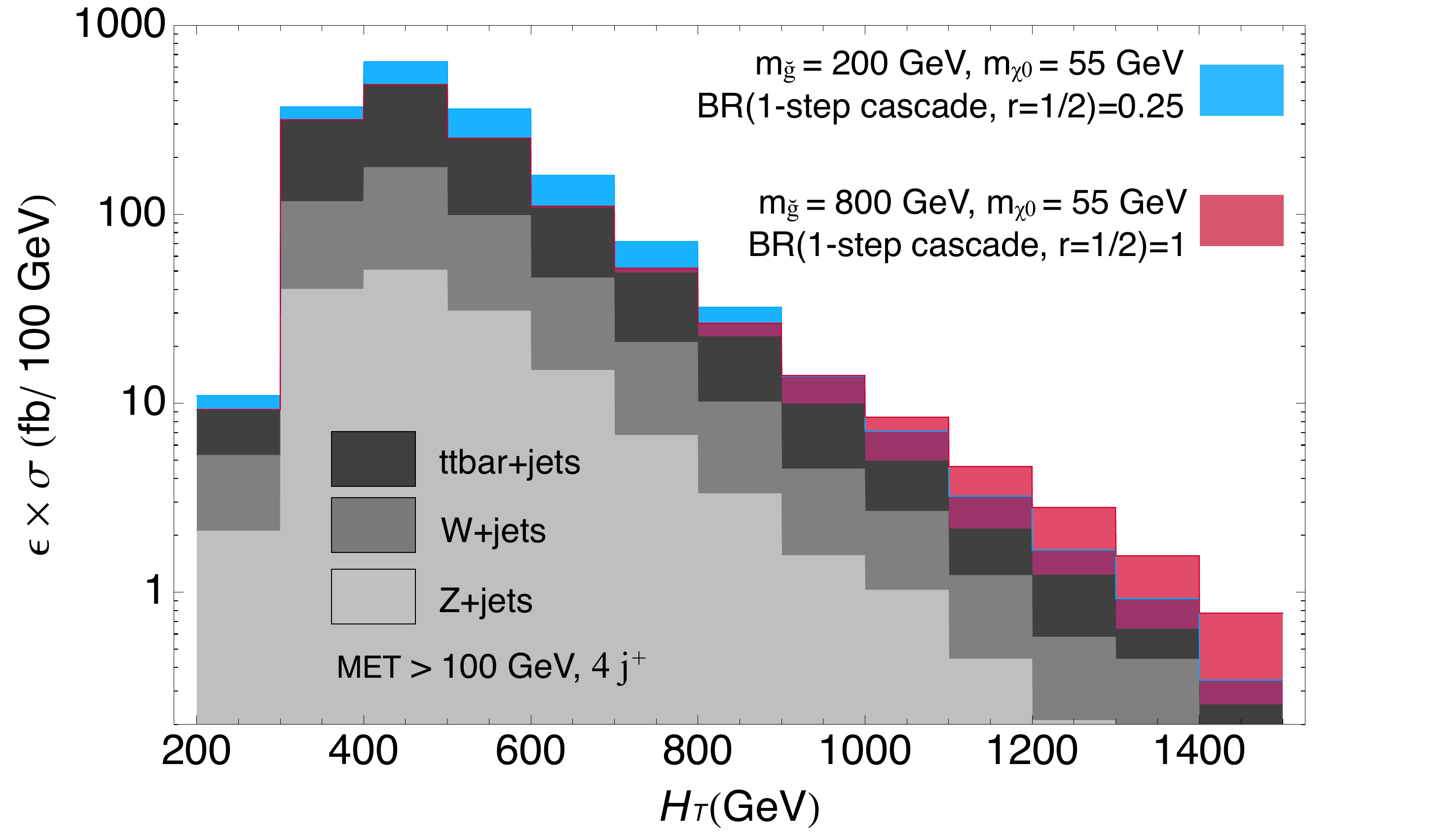}
 \end{center}
 \caption{
  \label{HThisto}
 $H_T$ distribution in the multijet channel with a $\MET>100\GeV$ requirement. The dominant backgrounds, $t\bar{t}+\text{jets}$ and $W/Z+\text{jets}$, are displayed in gray, whereas to signal points are displayed in color. The  Multijet Low MET search region, with a $H_T\gsim 450\GeV$ cut, would be sensitive to the 200 GeV $\go$ signal, whereas the Multijet High $H_T$ search region would be sensitive to the 800 GeV $\go$ signal.}
 \end{figure}

This {\it Multiregion Search Strategy} is crucial to ensure that new physics signatures will not be missed because there have been relatively few studies of the full signature space.  Performing a measurement in a single search region limits the reach to a preferred corner of model space. Only the combination of complementary search regions captures the full potential of the LHC, providing near-optimal coverage.

Fully studying any simplified model requires extensive Monte Carlo calculations of the different signals.  The primary concern of this article is to ensure that no signal is missed and to motivate more thorough searches in jets and missing energy.   Simulation of hypothetical signals is computationally costly and minimizing Monte Carlo generation is desirable, particularly for the experiments at the LHC where full detector simulations limit the size of surveys of hypothetical signals.  The search regions found here need to be reanalyzed by the experimentalists to make sure that they do provide the coverage claimed and are not limited by unforeseen backgrounds.   In order to facilitate this process, a selection of a few dozen benchmark simplified models is provided to ensure that sensitivity is not lost when tuning searches.  These benchmark simplified models are fully specified in App.~\ref{App: Benchmarks}, Tabs.~\ref{BM0.1}-\ref{BM2.1} and marked as dark dots in Fig.~\ref{MinCuts}.

\subsection{Alternative $p_T$ Selection Criteria}
The multiregion search strategy in $\MET$ and $H_T$ presented in the last subsection used the pre-selection criteria described in Sec.~\ref{Sec: Results}.  These pre-selection criteria are determined primarily by triggering and can in principle be tightened to provide better limits.   For the three-body direct decay, two-step cascade decay and $r=\frac{1}{4},\frac{1}{2}$ one-step cascade decay modes,  30\% to 40\% sensitivity can be gained in the range $350\GeV\lsim m_{\go}\lsim600\GeV$, $m_{\LSP}\lsim150\GeV$ by hardening the $p_T$-selection criteria on the sub-leading jets of the inclusive tetrajet channel:
\begin{equation}
 p_{T1},~ p_{T2},~ p_{T3},~ p_{T4}\geq100\GeV.
\end{equation}
These selection criteria could replace the ones in Sec.~\ref{Sec: Results}, with a gain in sensitivity in the low $m_{\LSP}$ region for some decay topologies. However, this tighter pre-selection results in a loss in sensitivity ranging between 20\% to 40\% in the whole parameter space for the $r=\frac{3}{4}$ one-step cascade decay mode.  It also renders the tetrajet channel inefficient for the two-body direct decay mode.  Tightening the pre-selection criteria in  lower multiplicity channels  to $p_{T}\geq100\GeV$ results in a 10\% to 40\% loss of sensitivity for more compressed spectra and would universally affect the reach for $m_{\go}-m_{\LSP}\lsim300\GeV$. 
This discussion shows that the search regions defined in the previous subsection can be improved, but at the cost of  a significantly more complicated design.

\section{Multiple Decay Modes}
\label{Sec: MultipleDecayModes}

The studies considered so far are applicable only when a single $\go$ decay mode contributes to the all-hadronic jets plus missing energy channel.  When the $\go$ has two or more decay modes contributing to this channel, translating results from the single $\go$ decay mode into these more generic decay patterns is not completely straightforward. 
This section addresses how to infer sensitivities of searches to simplified models when there are multiple decay modes, but only single decay modes have been explicitly studied.  Understanding how to use models with single decay modes in a more general context greatly enhances the applicability of simplified models and reduces the overall complexity of studying more general examples.

In the following we provide a quantitative illustration of an estimation of the sensitivity to the cross section for models with multiple $\go$ decay modes. 
Consider a model in which  $\go$ decays through two modes, $A$ and $B$, with branching ratios $\BB_A$ and $\BB_B$, respectively.   The goal is to be able to set a limit on the $\go$ pair-production cross section with these two decay modes,
\begin{eqnarray*} 
\sigma_{\text{prod}}(\BB_A,\BB_B|\mathcal{S}),
\end{eqnarray*}
when only the quantities
\begin{eqnarray*}
\sigma_{\text{prod}}(1,0|\mathcal{S}), 
\qquad
\sigma_{\text{prod}}(0,1|\mathcal{S})
\end{eqnarray*}
are known.

A conservative limit on the production cross-section can be extracted by considering the most constrained decay mode and ignoring the other decay modes,
\begin{eqnarray}
\sigma_\text{prod}(\BB_A,\BB_B)
\le \text{min } \left\{\frac{
\sigma_{\text{prod}}(1,0)}{\BB_A^2}, 
\frac{\sigma_{\text{prod}}(0,1)}{\BB_B^2}
\right\} .
\end{eqnarray}
This can be a considerable underestimate of the actual search sensitivities because it assumes that the mixed and less sensitive decay modes make no contribution to the signal region.

The sensitivity of a given search region $\mathcal{S}$ depends on the number of events of each topology that pass the cuts defining $\mathcal{S}$. The topologies $AA$, $BB$ and $AB$ contribute the following number of events to the signal region
\begin{eqnarray}
\label{modes}
\nonumber N_{AA}(\mathcal{S})=& \epsilon_{AA}(\mathcal{S}) \BB_A^2\times N
\qquad&\text{for}\quad \go\go\rightarrow AA,\\
N_{BB}(\mathcal{S})=&\epsilon_{BB}(\mathcal{S})\BB_B^2\times N
\qquad&\text{for}\quad \go\go\rightarrow BB,\\
\nonumber N_{AB}(\mathcal{S})=&2\epsilon_{AB}(\mathcal{S})\BB_A\BB_B\times N
\qquad&\text{for}\quad \go\go\rightarrow AB,
\end{eqnarray}
where  $\epsilon(\mathcal{S})$ is the efficiency for the given decay mode to pass the cuts defining the search region $\mathcal{S}$ and 
\begin{eqnarray*}
N(\mathcal{S})= \LL \times \sigma_\text{prod}(\BB_A,\BB_B|\mathcal{S})
\end{eqnarray*}
is the total number of events.  It is useful to define the weighted efficiency for $\mathcal{S}$ 
\begin{eqnarray}
\epsilon_T(\BB_A, \BB_B| \mathcal{S}) = \epsilon_{AA}(\mathcal{S})\BB_A^2 + \epsilon_{BB}(\mathcal{S})\BB_B^2+2\epsilon_{AB}(\mathcal{S})\BB_A\BB_B .
\end{eqnarray}

\begin{figure}[h]
\begin{center}
\includegraphics[width=5in]{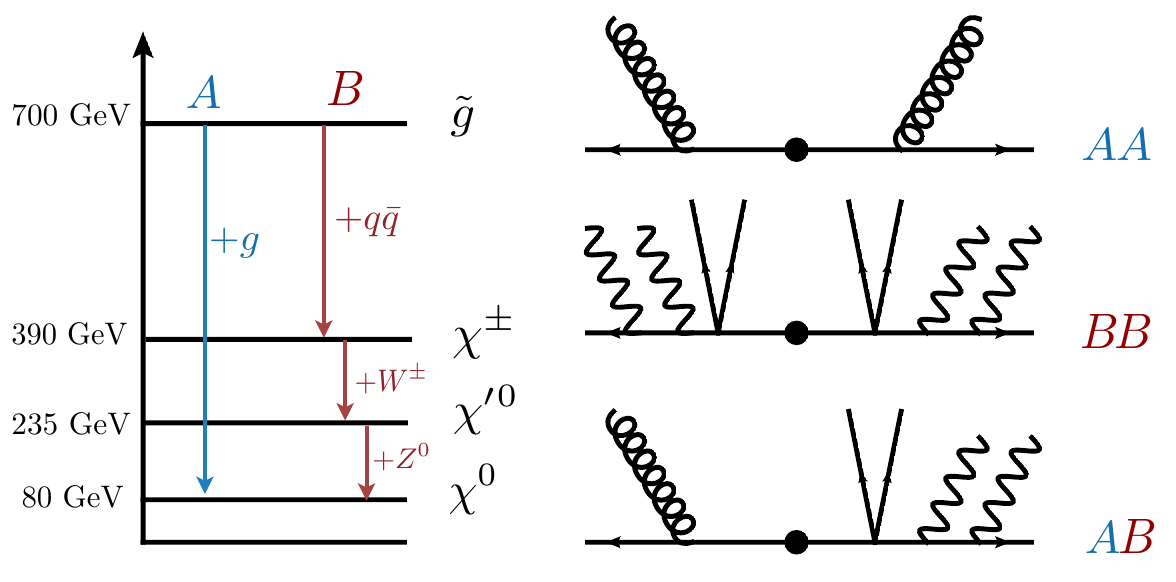}
 \end{center}
 \caption{
 \label{Fig: HybridFig}
 Benchmark model $m_{\go}=700~\GeV$, $m_{\LSP}=80~\GeV$ with two $\go$ decay modes: $A=$ two-body direct decay 
 and $B=$ two-step cascade decay with $r=r'=\half$. Three topologies are possible: non-hybrid $AA$, $BB$ and hybrid $AB$.  This example
 is studied in the text with $\BB_{A}=\BB_B=50\%$.
 }
 \end{figure}

As a specific example, Fig.~\ref{Fig: HybridFig} illustrates a theory 
with benchmark masses $m_{\go}=700~\text{GeV},~ m_{\LSP}=80~\text{GeV}$ in the case in which mode $A$ corresponds to the 2-body direct decay of $\go$ and mode $B$ to the 2-step cascade decay.
Tab.~\ref{hybrid} displays the efficiencies and background estimates for the search regions of Sec.~\ref{MSRS}.

The expected $2\sigma$-sensitivity to the cross section is extracted by demanding that the number of signal events in $\mathcal{S}$, $N(\mathcal{S})=\epsilon_T(\mathcal{S}) \times N$, be greater than twice the background uncertainty.
From \eqref{Eq: SigmaLimit}, the 2$\sigma$ limit on the production cross section from $\mathcal{S}$ is
\begin{eqnarray}
\sigma^{2\sigma}_\text{prod}(\BB_A,\BB_B|\mathcal{S})
=
\frac{\Delta B^{2\sigma} (\mathcal{S})}{\LL\times\epsilon_T(\BB_A, \BB_B, \mathcal{S})}
\label{computeXsection}
\end{eqnarray}
where $\Delta B^{2\sigma}$ is the $2\sigma$ uncertainty in the background given in  \eqref{Eq: MinEvents}.

The question of how to extract limits for theories with multiple decay modes reduces to how much is known about $\epsilon_{AB}(\mathcal{S})$ when only $\epsilon_{AA}(\mathcal{S})$ and $\epsilon_{BB}(\mathcal{S})$ are known.
Although it is possible to obtain the efficiencies for hybrid events through separate Monte Carlo studies, when there are several different decay modes it is computationally expensive to extract the efficiencies for all hybrid events ($\half N_d(N_d+1)$ additional Monte Carlo studies when $N_d$ different decays are present).  A more conservative approach is to estimate the efficiencies for hybrid events using the known efficiencies for non-hybrid events. 
If the decay modes $A$ and $B$ are relatively similar, then the efficiencies for the hybrid
events are bounded by
\begin{eqnarray}
\text{min}\left\{ \epsilon_{AA}(\mathcal{S}),\epsilon_{BB}(\mathcal{S}) \right\}
\le \epsilon_{AB}(\mathcal{S}) \le 
\text{max}\left\{ \epsilon_{AA}(\mathcal{S}),\epsilon_{BB}(\mathcal{S}) \right\} .
\end{eqnarray}
When this two-sided bound is satisfied, $\sigma_\text{prod}(\go\go X;\BB_A,\BB_B|\mathcal{S})$ can be bounded from both sides 
\begin{eqnarray}
\frac{\sigma^{2\sigma}_\text{prod}(1,0|\mathcal{S})}{ \BB_A^2 +(\BB_B^2 + 2\BB_B\BB_A )\frac{ \epsilon_{BB}}{\epsilon_{AA}} } 
\le 
\sigma^{2\sigma}_\text{prod}(\BB_A,\BB_B|\mathcal{S})
\le
\frac{\sigma^{2\sigma}_\text{prod}(1,0|\mathcal{S})}{ \BB_A^2 +2 \BB_A \BB_B + \BB_B^2 \frac{ \epsilon_{BB}}{\epsilon_{AA}}}  
\end{eqnarray}
assuming $\epsilon_{AA} \le \epsilon_{BB}$.

This two-sided bound on efficiencies for hybrid decays is not always satisfied when the decay modes are substantially different, {\it e.g.} two-body direct decay {\it vs.} two-step cascade decay.  In general there is no way to bound $\epsilon_{AB}$ in terms of $\epsilon_{AA}$ and $\epsilon_{BB}$.
However, for the search regions studied in this article, the efficiencies for the hybrid decays do satisfy a  bound of
\begin{equation}
\label{epsAB}
\epsilon_{AB}(\mathcal{S})\geq\text{min}\left\{\epsilon_{AA}(\mathcal{S}),\epsilon_{BB}(\mathcal{S})\right\},
\end{equation}
which was verified empirically.  By taking the lower bound of the relation above, a conservative estimation of the cross section sensitivity can be placed,
\begin{eqnarray}
\frac{ \sigma^{2\sigma}_\text{prod}(1,0|\mathcal{S})}{ \BB_A^2 + 2\BB_B\BB_A\frac{1}{\epsilon_{AA}} + \BB_B^2\frac{ \epsilon_{BB}}{\epsilon_{AA}}} 
\le 
\sigma^{2\sigma}_\text{prod}(\BB_A,\BB_B|\mathcal{S})
\le
\frac{\sigma^{2\sigma}_\text{prod}(1,0|\mathcal{S})}{ \BB_A^2 +2 \BB_A \BB_B + \BB_B^2 \frac{ \epsilon_{BB}}{\epsilon_{AA}}}  
\end{eqnarray}
assuming that $\epsilon_{AA}\le \epsilon_{BB}$. The lower limit on $\sigma^{2\sigma}_\text{prod}(\BB_A,\BB_B|\mathcal{S})$ arises from the constraint $\epsilon_{AB}\le 1$ and is primarily useful when the $AB$ topology branching ratio is subdominant to the $AA$ or $BB$ branching ratio.
 In Tab.~\ref{guess} this conservative estimate of the cross section is applied to the specific example of Fig.~\ref{Fig: HybridFig}, with $\BB_A=\BB_B=50\%$.
 Notice that even though the search regions were not explicitly designed to be sensitive to these hybrid decays, sensitivity is nearly optimal with $\EE = 1.02$ for the most sensitive search region. 
The optimal sensitivity for this benchmark simplified model is set by the high-$H_T$ cut of the multiple search region. The conservative estimate for the limit on the production cross-section (162~fb) is higher than the actual value (108~fb) by $\approx 49\%$, and higher than the {\it optimal} value (106~fb) by $\approx 53\%$.  More generally, in the framework of comprehensive multiregion search strategies, this conservative estimate on the cross section limit is within 50\% of the actual limit for all of the simplified models studied.

\begin{table}[htdp]
\begin{center}
\begin{tabular}{|c||c|c|c||c|c||c|c|c|}
\hline
$~\mathcal{S}~$ & Ch & $\MET$(GeV) & $H_T$(GeV) & $\sigma_\text{bg}\times \epsilon$ (fb) & $\Delta B^{2\sigma}$ & $\epsilon_{AA}$ & $\epsilon_{BB}$ & $\epsilon_{AB}$\\
\hline\hline
1& $2^+j$ & 500 & 750 & 14.1 & 11.2& 0.11 & 0.0017 & 0.028 \\
\hline
\colorbox{white}{2}& \colorbox{white}{$3^+j$} & \colorbox{white}{450} & \colorbox{white}{500 }& \colorbox{white}{22.1} &\colorbox{white}{ 16.2} &\colorbox{white}{ 0.17} & \colorbox{white}{0.0038} &\colorbox{white}{ 0.063}\\
\hline
3& $4^+j$ & 100 & 450 & 667.7 & 404 & 0.25 & 0.20 & 0.43 \\
\hline
4& $4^+j$ & 150 & 950 & 13.4 & 10.4 &  0.075 & 0.064 & 0.13 \\
\hline
5& $4^+j$ & 250 & 300 & 105.7 & 66.6 & 0.21 & 0.078 & 0.33 \\
\hline
6& $4^+j$ & 350 & 600 & 18.1 & 13.8 & 0.14 & 0.017 & 0.15 \\
\hline
\end{tabular}
\caption{
\label{hybrid} 
Signal efficiencies for the multiple search region of Sec.~\ref{MSRS} for benchmark masses $m_{\go}=700\text{GeV},~ m_{\LSP}=80\text{GeV}$ and $\go$ decay modes $A=$ 2-body direct decay and $B=$ 2-step cascade decay. Also included are the expected background cross section in the signal region $\sigma_\text{bkg}\times \epsilon$ and corresponding statistical and systematic uncertainties $\Delta B^{2\sigma}$ for a luminosity of $1~\text{fb}^{-1}$.
}
\end{center}
\end{table}

\begin{table}[htdp]
\begin{center}
\begin{tabular}{|c||c|c||c|c|c||c|c|}
\hline
$~\mathcal{S}~$ &  $\sigma_{\text{prod}}^{2\sigma}(1,0)$& $\sigma_{\text{prod}}^{2\sigma}(0,1)$ & $\sigma^{2\sigma}_\text{actual}(0.5,0.5)$ & $\sigma^{2\sigma}_\text{cons.} (0.5,0.5)$ & $\sigma^{2\sigma}_\text{cons.}/\sigma^{2\sigma}_\text{actual}$ & $\mathcal{E}_{\text{actual}}$& $\mathcal{E}_{\text{cons.}}$\\
\hline\hline
1& 101 fb&6690 fb& 267   fb& 389  fb & 1.46 & 2.52&3.67  \\
\hline
\colorbox{orange}{2}&\colorbox{orange}{95.3 fb}&\colorbox{orange}{4260 fb } & \colorbox{orange}{216 fb}  &\colorbox{orange}{ 357 fb}  &\colorbox{orange}{ 1.65 }& \colorbox{orange}{2.04 }&\colorbox{orange}{3.37 }  \\
\hline
3& 1610 fb&2020 fb & 1234   fb& 1901 fb  & 1.54 &11.6 &17.9 \\
\hline
\colorbox{yellow}{4}&\colorbox{yellow}{144 fb}&\colorbox{yellow}{169 fb} & \colorbox{yellow}{108  fb} & \colorbox{yellow}{162 fb } &  \colorbox{yellow}{1.49} & \colorbox{yellow}{1.02}& \colorbox{yellow}{1.53}  \\
\hline
5& 317 fb&854  fb& 281  fb & 600  fb & 2.13 &2.65 & 5.66 \\
\hline
6& 98.6 fb&812 fb& 121 fb  & 289 fb  & 2.39 & 1.14 &2.73 \\
\hline
\end{tabular}
\caption{
\label{guess} Cross section sensitivity for the benchmark masses $m_{\go}=700\text{GeV},~ m_{\LSP}=80\text{GeV}$ and the two $\go$ decay modes $A=$ 2-body direct decay and $B=$ 2-step cascade decay, where, in order to maximize the number of hybrid events, we take $\BB_A=\BB_B=50\%$. 
The highlighted orange search is most effective for the $AA$ topology and the yellow is most sensitive for the $BB$ and $AB$ topologies. The actual sensitivity $\sigma_\text{actual}$ was computed using \eqref{computeXsection} and the efficiencies displayed in Tab.~\ref{hybrid}. The conservative estimate $\sigma_\text{cons}$ in case the efficiency for hybrid events is unknown is obtained by taking the lower bound in \eqref{epsAB} for $\epsilon_{AB}$. The last column displays the efficacy of each search region under the conservative estimates $\sigma_\text{cons}$. The efficacy, $\EE$, is defined in \eqref{effic} and quantifies how close the cross section limits are from the optimal one, $\sigma_\text{optimal}=106$~fb.
}
\end{center}
\end{table}

\section{Discussion}
\label{discussion}

This work focused on the optimization of searches for new colored states with jets plus missing energy signatures. Simplified Models were used to capture relevant new physics features with the simplest spectra in the most important decay topologies.
With optimized search regions, the reach of the 7 TeV LHC for heavy colored octets decaying to jets and a long-lived invisible particle was estimated in a wide range of masses and the following decay topologies: two- and three-body direct decays, one-step cascade decays and two-step cascade decays. 

The optimal reach requires tuning the search regions for each individual spectrum, and is therefore not practical. A more minimal search strategy, nearly as effective as a fully optimized search, was presented, consisting of a set of six search regions. These search regions are characterized by cuts on missing and visible energy, whose combined reach is within 30\% of optimal for all kinematic regimes, decay topologies, and integrated luminosities in the range $10\ipb\leq \LL \leq 1\ifb$. Although not unique, multiregion search strategies share qualitative features that capture specific regions of the phase space of signatures. For instance, a hard missing transverse energy cut in the inclusive dijet channel is required for coverage of compressed spectra. Other regions are best covered by higher jet multiplicities, in particular the tetrajet channel. 

The efficacy of the search regions depends on the assumptions about the backgrounds, in particular the systematic uncertainty.  For instance, the studies in this article used a 30\% systematic uncertainty on all backgrounds.  If there are larger systematic uncertainties, search regions with harder cuts will be preferred so as to remove backgrounds, whereas looser cuts will be preferred for smaller systematics.  Since these uncertainties are moving targets, a set of benchmark simplified models are given in App.~\ref{App: Benchmarks} that provide a representative sampling of the whole space of models and can be used for designing a comprehensive multiregion search strategy. These benchmark models can be used to optimize search strategies with more realistic background calculations or background measurements.

The results of this work reveal a promising picture  for the upcoming LHC results that will shortly be released.  Specifically, with $45\ipb$, the LHC will be able to test $\go$ masses up to 600~GeV for light $\LSP$'s, and have nearly complete coverage, independent of $m_{\LSP}$, up to 350~GeV.    With $1\ifb$, the reach on $\go$ masses will extend up to 850~GeV and have complete coverage up to 400~GeV.  Jets and missing energy are the first channels to look for new physics with QCD interactions below the TeV scale, and multiregion search strategies will play a key role in the discovery process.

\vspace{0.5in}
%%%%%%%%%%%%%%%%%%%%%%%%%%%%%%%%%%%%%%%
\noindent
{ \bf Acknowledgements}

We would like to thank Amir Farbin, Louise Heelan, Zachary Marshall, Mariangela Lisanti and Josh Ruderman for helpful discussions. DSMA, EI and JGW are supported by the DOE under contract DE-AC03-76SF00515. JGW is partially supported by the DOE's Outstanding Junior Investigator Award.

\newpage
\appendix

\section{Reach Estimates}
\label{App: PlotHell}

\begin{figure}[h]
\begin{center}
\includegraphics[width=5.2in]{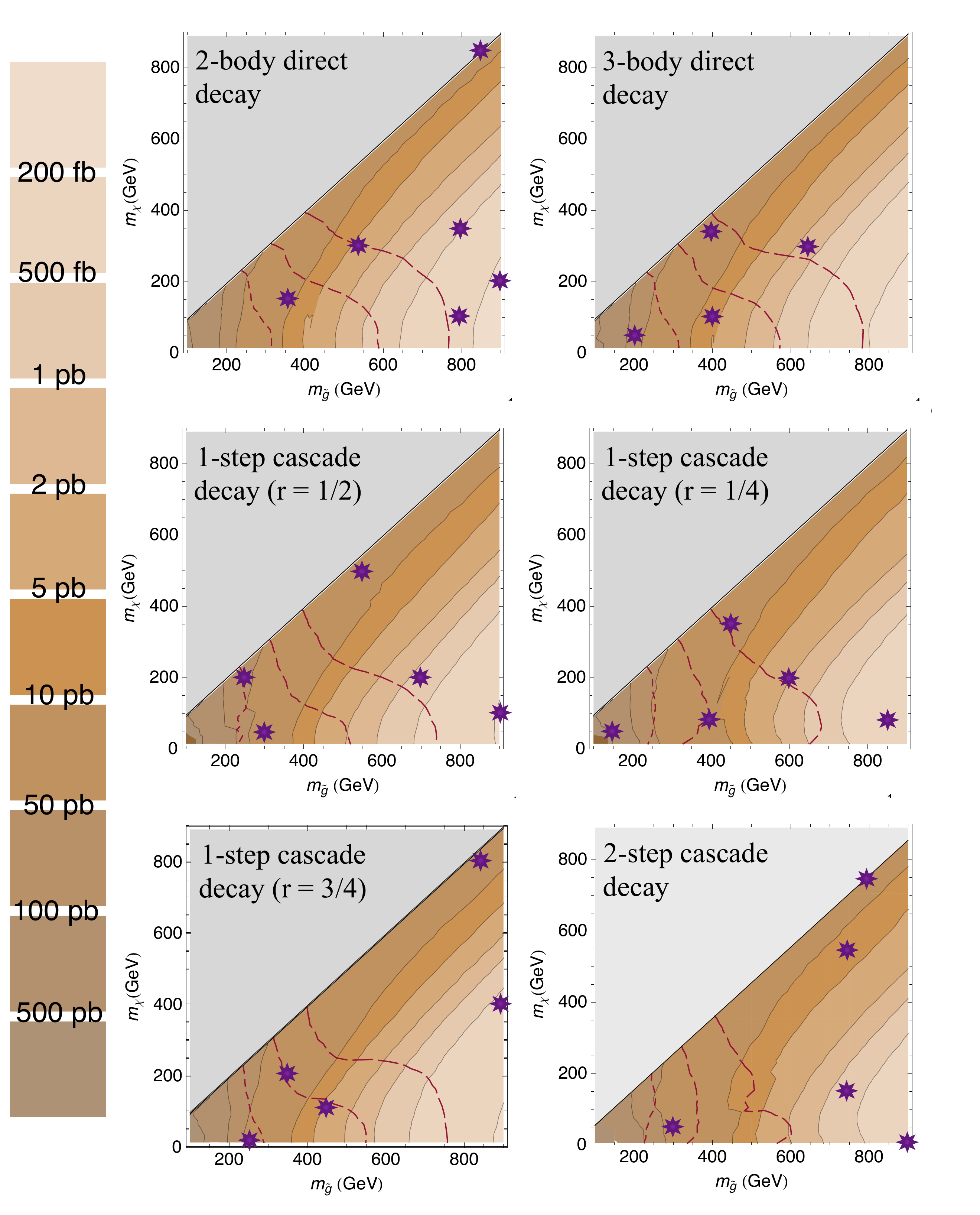}
 \end{center}
 \caption{
  \label{AllCrossSections45pb}
 Contours of the $2\sigma$-sensitivity for $\sigma(pp\rightarrow\go\go)\times\BB$ for the six decay modes considered in this work, assuming $\LL=45\ipb$ and $\sqrt{s}=7\TeV$.  The contour values are specified on the color scale on the left. The red (dashed) lines are simple parametrizations of $\sigma_\text{prod}\times\BB$ in terms of the $\go$ NLO-QCD production corresponding to $\sigma_{\text{prod}} \times\BB/\sigma_{\text{NLO-QCD}} = 3.0, 1.0, 0.3$, moving from right to left, respectively.
 }
 %. They correspond to: (i) $\sigma_\text{prod}\times\BB_{\go}=3\times\sigma_\text{NLO-QCD}$ (right-most long-dashed line), (ii) $\sigma_\text{prod}\times\BB_{\go}=\sigma_\text{NLO-QCD}$ (middle dashed line), and (iii) $\sigma_\text{prod}\times\BB_{\go}=0.3\times\sigma_\text{NLO-QCD}$ (left-most short-dashed line).}
 \end{figure}

\begin{figure}[h]
\begin{center}
\includegraphics[width=5.2in]{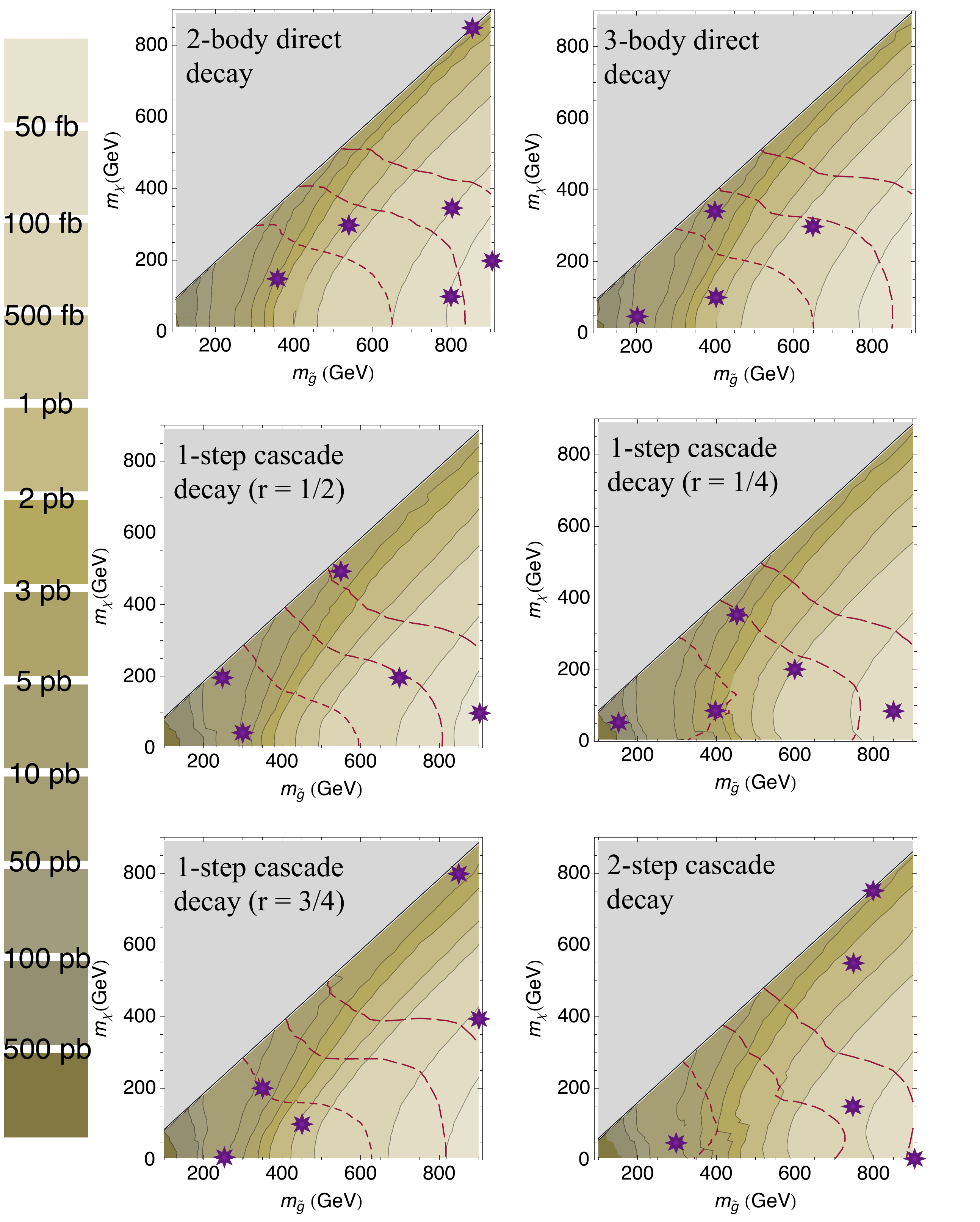}
 \end{center}
 \caption{
  \label{AllCrossSections}
 Same as in Fig.~\ref{AllCrossSections45pb} but with $1\ifb$ of integrated luminosity at $\sqrt{s}=7~\TeV$.}
 \vspace{0.01in}
 \end{figure}
 
 \newpage
 {\tiny .}\\
 \pagebreak[4]
 
\section{Benchmark Models}
\label{App:  Benchmarks}
This appendix lists the fully specified benchmark models.

\addtocounter{table}{-2}
\def\thetable       {3.\Alph{table}}

\begin{table}[h]
\begin{center}
\begin{tabular}{|c||c|c||c|c||c|}
\hline
Name & $m_{\tilde{g}}$ (GeV)& $m_{\tilde{\chi}^0}$ (GeV)  & $\sigma^\text{opt}_{45 \pb^{-1}}$ (pb) & $\sigma^\text{opt}_{1 \fb^{-1}}$ (pb)&$\sigma^{\text{\tiny{QCD}}}_{\text{prod}}$ (pb) \\
\hline\hline
$\GG^{\tt  2DD}_1$& 350 & 150 & 13  & 2.9 & 24 \\
\hline
$\GG^{\tt 2DD}_2$& 550 & 300 & 3.2 & 0.69 & 1.2  \\
\hline
$\GG^{\tt 2DD}_3$& 800 & 100 & 0.23 & 0.049  & 0.061 \\
\hline
$\GG^{\tt 2DD}_4$& 800 & 350 &0.46& 0.097 & 0.061\\
\hline
$\GG^{\tt  2DD}_5$& 850 & 840 & 14& 3.3 & 0.036 \\
\hline
$\GG^{\tt 2DD}_6$& 900 & 190 & 0.17&0.036   & 0.021 \\
\hline
\end{tabular}
\caption{\label{BM0.1} Benchmark simplified models for $\go$'s that decay directly to $\chi^0$ through a two-body decay. The optimal reach in the cross section for each one of the benchmark models, $\sigma^\text{opt}$, is displayed for two luminosity scenarios: $45\ipb$ and $1\ifb$. Also displayed is the reference NLO-QCD cross section for $\go$ pair-production.  }
\end{center}
\end{table}
\begin{table}[h]
\begin{center}
\begin{tabular}{|c||c|c||c|c||c|}
\hline
Name & $m_{\tilde{g}}$ (GeV)& $m_{\tilde{\chi}^0}$ (GeV)  & $\sigma^\text{opt}_{45 \pb^{-1}}$ (pb) & $\sigma^\text{opt}_{1 \fb^{-1}}$ (pb)&$\sigma^{\text{\tiny{QCD}}}_{\text{prod}}$ (pb) \\
\hline\hline
$\GG^{\tt 3DD}_1$& 250& 50 &32  &9.5   & 180  \\
\hline
$\GG^{\tt 3DD}_2$& 400 & 100 & 5.7 &1.2  & 14 \\
\hline
$\GG^{\tt 3DD}_3$& 400 & 350 &23  &5.9 & 14 \\
\hline
$\GG^{\tt 3DD}_{4}$& 650 & 300 & 1.2 & 0.26 & 0.34\\
\hline
\end{tabular}
\caption{\label{BM0.2} Benchmark simplified models for $\go$'s that decay directly to $\chi^0$ through a three-body decay.   }
\end{center}
\end{table}
\begin{table}[h]
\begin{center}
\begin{tabular}{|c||c|c|c||c|c||c|}
\hline
Name & $m_{\tilde{g}}$ (GeV)&$m_{\chi^\pm}$ (GeV) &$m_{\chi^0}$ (GeV)  & $\sigma^\text{opt}_{45 \pb^{-1}}$ (pb) & $\sigma^\text{opt}_{1 \fb^{-1}}$ (pb) &
$\sigma^{\text{\tiny{QCD}}}_{\text{prod}}$ (pb)  \\
\hline\hline
$\GG^{\tt 1CD}_{1}$& 150 & 75&50 &214 & 156 & 2900\\
\hline
$\GG^{\tt 1CD}_{2}$& 400&160 & 80 &10  &  2.3 & 14\\
\hline
$\GG^{\tt 1CD}_{3}$& 450 & 375&350  &17  & 4.2 & 4.8\\
\hline
$\GG^{\tt 1CD}_{4}$& 600 &300 &200 & 2.2 & 0.48  & 0.62 \\
\hline
$\GG^{\tt 1CD}_{5}$& 850 & 272.5 &80  & 0.30& 0.064 & 0.036\\
\hline
\end{tabular}
\caption{\label{BM1.1} Benchmark simplified models for $\go$'s that decay through a one-step cascade to $\chi^0$ with $r=1/4$, where $r= (m_{\chi^\pm}- m_{\chi^0})/(m_{\tilde{g}} - m_{\chi^0})$.  }
\end{center}
\end{table}
\begin{table*}[h]
\begin{center}
\begin{tabular}{|c||c|c|c||c|c||c|}
\hline
Name & $m_{\tilde{g}}$ (GeV)&$m_{\chi^\pm}$ (GeV) &$m_{\chi^0}$ (GeV)  & $\sigma^\text{opt}_{45 \pb^{-1}}$ (pb) & $\sigma^\text{opt}_{1 \fb^{-1}}$  (pb) &
$\sigma^{\text{\tiny{QCD}}}_{\text{prod}}$ (pb)  \\
\hline
\hline
$\GG^{\tt 1CD}_{6}$& 250 & 225 &200  & 57 &23 & 180\\
\hline
$\GG^{\tt 1CD}_{7}$& 300 & 175&50  &27 &7.4 & 62 \\
\hline
$\GG^{\tt 1CD}_{8}$& 550 & 525 &500 & 18 & 4.5 & 1.2 \\
\hline
$\GG^{\tt 1CD}_{9}$& 700 & 550&200  & 0.84 &  0.18 & 0.19 \\
\hline
$\GG^{\tt 1CD}_{10}$& 900 & 500 & 100  &  0.20 & 0.042 & 0.021\\
\hline
\end{tabular}
\caption{\label{BM1.2} Benchmark simplified models for $\go$'s that decay through a one-step cascade to $\chi^0$ with $r=1/2$.  }
\end{center}
\end{table*}

\begin{table*}[h]
\begin{center}
\begin{tabular}{|c||c|c|c||c|c||c|}
\hline
Name & $m_{\tilde{g}}$ (GeV)&$m_{\chi^\pm}$ (GeV) &$m_{\chi^0}$ (GeV)  & $\sigma^\text{opt}_{45 \pb^{-1}}$ (pb) & $\sigma^\text{opt}_{1 \fb^{-1}}$ (pb) &
$\sigma^{\text{\tiny{QCD}}}_{\text{prod}}$ (pb)  \\
\hline\hline
$\GG^{\tt 1CD}_{11}$& 250 &187.5 &0  & 37 & 12 & 180\\
\hline
$\GG^{\tt 1CD}_{12}$& 350 &312.5 &200  &27 &  7.5  & 24 \\
\hline
$\GG^{\tt 1CD}_{13}$& 450 &362.5 &100 & 4.4 &0.94 & 4.8 \\
\hline
$\GG^{\tt 1CD}_{14}$& 850 & 837.5 & 800 &15  & 3.4 & 0.036 \\
\hline
$\GG^{\tt 1CD}_{15}$& 900 & 775&400  & 0.42 & 0.090 & 0.021\\
\hline
\end{tabular}
\caption{\label{BM1.3} Benchmark simplified models for $\go$'s that decay through a one-step cascade to $\chi^0$ with $r=3/4$. }
\end{center}
\end{table*}
\begin{table}[b]
\begin{center}
\begin{tabular}{|c||c|c|c|c|c|c|c|c|}
\hline
Name & $m_{\tilde{g}}$ (GeV)& $m_{\chi^\pm}$ (GeV)& $m_{\chi'{}^0}$ (GeV)& $m_{\chi^0}$ (GeV) &  $\sigma^\text{opt}_{45 \pb^{-1}}$ (pb) & $\sigma^\text{opt}_{1 \fb^{-1}}$ (pb) & $\sigma^{\text{\tiny{QCD}}}_{\text{prod}}$ (pb)\\
\hline\hline
$\GG^{\tt 2CD}_{1}$& 300&175&112.5 & 50 & 37  & 12    & 62 \\
\hline
$\GG^{\tt 2CD}_{2}$& 750 &450 &300&150 &  0.67  &  0.14  & 0.11\\
\hline
$\GG^{\tt 2CD}_{3}$& 750 &650 &600&550 &   6.6   & 1.4  & 0.11\\
\hline
$\GG^{\tt 2CD}_{4}$& 800 & 775&762.5&750 &  15  &     3.4 & 0.061\\
\hline
$\GG^{\tt 2CD}_{5}$& 900 &  450 & 225 & 0 &   0.33 &    0.070 & 0.021\\
\hline
\end{tabular}
\caption{\label{BM2.1} Benchmark simplified models for $\go$'s that decay through a 2-step cascade to $\chi^0$ with $r=r'=1/2$, where $r= (m_{\chi^\pm}- m_{\chi^0})/(m_{\tilde{g}} - m_{\chi^0})$ and $r'=(m_{\chi'{}^0}- m_{\chi^0})/(m_{\chi^\pm} - m_{\chi^0})$. }
\end{center}
\end{table}

\newpage
{\tiny .}
%%%%%%%%%%%%%%%%%%%%%%%%%%%%%
\pagebreak[4]

\end{document}